\documentclass[aps,prmaterials,10pt,twocolumn,amsmath,amssymb,amsfonts,longbibliography,superscriptaddress,nofootinbib]{revtex4-2}
\usepackage{graphicx}
\usepackage{hhline}
\usepackage[latin1]{inputenc}
\usepackage{subfigure}

\usepackage{xcolor}

\begin{document}

\title{Influence of static correlation on the magnon dynamics of an itinerant ferromagnet with competing exchange interactions - a first principles study of MnBi}

\author{Thorbj\o rn Skovhus}
\affiliation{CAMD, Department of Physics, Technical University of Denmark, 2800 Kgs. Lyngby, Denmark}
\author{Thomas Olsen}
\email{tolsen@fysik.dtu.dk}
\affiliation{CAMD, Department of Physics, Technical University of Denmark, 2800 Kgs. Lyngby, Denmark}
\author{Henrik M. R\o nnow}
\affiliation{LQM, Institute of Physics, \'{E}cole Polytechnique F\'{e}d\'{e}rale de Lausanne, CH-1015 Lausanne, Switzerland}

\begin{abstract}
We present first principles calculations of the dynamic susceptibility in strained and doped ferromagnetic MnBi using time-dependent density functional theory. In spite of being a metal, MnBi exhibits signatures of strong correlation and a proper description in the framework of density functional theory requires Hubbard corrections to the Mn $d$-orbitals. To permit calculations of the dynamic susceptibility with Hubbard corrections applied to the ground state electronic structure, we use a consistent rescaling of the exchange-correlation kernel maintaining the delicate balance between the magnon dispersion and the Stoner continuum. We find excellent agreement with the experimentally observed magnon dispersion for pristine MnBi and show that the material undergoes a phase transition to helical order under application of either doping or strain. The presented methodology
paves the way for future LR-TDDFT studies of magnetic phase transitions, also for the wide range of materials with pronounced static correlation effects that are not accounted for at the LDA level. 
\end{abstract}

\maketitle

\section{Introduction}
MnBi has been proposed as a promising alternative to permanent magnets based on rare-earth elements \cite{Adams1952,Yang2001,Ly2015}. This is primarily due to the large spontaneous magnetization at room temperature \cite{Roberts1956}, strong uniaxial magnetic anisotropy \cite{Williams1957} and the abundance of its constituent elements \cite{Coey2011}. Moreover, it exhibits an extraordinarily large Kerr rotation \cite{Di1992}, which makes it an ideal candidate for magnetooptical data storage applications \cite{Williams1957a, Chen1973}. From a fundamental point of view MnBi is an 
intriguing material since it can be regarded as a member of the transition metal pnictides AX (A=V/Cr/Mn/Fe/Co/Ni, X=P/As/Sb), which realize a wide variety of spiral magnetic orders depending on the composition, pressure and temperature \cite{Wang2016, Goodenough1964, Goodenough1967}. In particular, MnP and CrAs have recently been demonstrated to exhibit unconventional superconductivity below 1 K \cite{Cheng2015} and 2 K \cite{Wu2014} respectively. The superconducting state emerges at the critical pressure for magnetic order and has been argued to imply an unconventional mechanism for Cooper pairing associated with magnetic quantum fluctuations \cite{Wu2014}. 
Thus, the microscopic origin of different spiral orders in transition metal pnictides is of high interest along with the influence of doping and strain. 
For example, at low temperatures MnP exhibits helical order at ambient pressure, becomes ferromagnetic at $\sim 1$ GPa, makes a transition to a second helical phase at $\sim 1.5$ GPa and finally becomes paramagnetic at 6.7 GPa with a superconducting window at pressures $6.7<P<8$ GPa \cite{Wang2016}. While MnBi has not been reported to exhibit phase transitions involving helical order, a phase transition from ferromagnetic to helical order has has been observed for the isostructural MnAs upon doping with Cr \cite{Luo2017} and it is not unlikely that a similar transition may occur for MnBi.

At room temperature, MnBi crystallizes in the hexagonal NiAs structure with space group $P6_3/mmc$. 
Upon cooling, MnBi undergoes a structural phase transition at 90 K, where the in-plane hexagonal symmetry is slightly broken and the crystal structure becomes orthorhombic (space group $Cmcm$) \cite{McGuire2014}. It is ferromagnetic up to 630 K where is segregates into Mn$_{1.08}$Bi and Bi \cite{Chen1974}, which are both paramagnetic. The true Curie temperature is thus unknown, but the lower bound of 630 K indicates strong magnetic interactions in the material. At room temperature, MnBi exhibits a large uniaxial magnetic anisotropy along the $c$-axis of the NiAs crystal structure. The anisotropy decreases with decreasing temperature and aligns with the $ab$-plane of the orthorhombic phase below 90 K \cite{McGuire2014}. For thin film \cite{Bandaru1998,Bandaru1999} and nanocrystalline \cite{Anand2019} MnBi, it has been shown that Cr doping lowers the Curie temperature below the segregation temperature for Mn$_{1-x}$Cr$_{x}$Bi doping levels down to at least $x=0.03$. For Cr doping in the range of 3-15\%, Curie temperatures have been recorded to lie in the 520-560 K range. Moreover, the coercivity has been shown increase significantly by Cr doping \cite{Anand2019}, while the beneficial magneto-optical properties are retained. 
To our knowledge there has so far not been attempts to study the low-temperature magnetic structure of Cr-doped MnBi.

The magnon spectrum of MnBi was been measured at 5 K by Williams et al. using inelastic neutron scattering (INS) \cite{Williams2016}. The hexagonal phase was assumed in the analysis of results, which is likely to be a good approximation due to the strong similarity with the orthorhombic phase expected at low temperatures \cite{Williams2016}. The magnon band width is on the order of 100 meV and the spectrum was found to be gapless as expected from the easy-plane anisotropy at low temperatures. It was shown that the spectrum is well reproduced by fitting to an isotropic Heisenberg model with 6 exchange parameters. While such a fit may not be unique, the nearest neighbor interaction was unambiguously shown to be antiferromagnetic having the largest magnitude of all interactions. The inherent magnetic frustration thus makes it highly plausible that a magnetic phase transition may be induced by external strain or doping, which will inevitable influence the individual exchange parameters differently.

From a computational perspective, the magnon spectrum of magnetic materials can be accessed at various levels of theory. Typically, a theoretical treatment will take one of the following three starting points: A mapping to a model Hamiltonian, an adiabatic approximation separating the transverse magnetic fluctuations from the faster electronic (longitudinal) degrees of freedom or an \textit{ab initio} treatment of the full dynamic transverse magnetic susceptibility. In the following, these three starting points will be referred to as model-based approaches, adiabatic approaches and theoretical magnon spectroscopy respectively. However, this classification is not unique and approaches with different starting points may in practise end up in equivalent overall treatments. 

The simplest model-based approach is to assume a Heisenberg model and extract exchange parameters in a total energy mapping analysis within the framework of density functional theory (DFT) \cite{Xiang2013,Olsen2017,Torelli2020}. This will directly yield the spin wave energy dispersion in the noninteracting magnon approximation to the model, but will not elucidate itinerant electron effects such as the lineshape broadening (Landau damping) from the spectral overlap with the single particle excitation continuum in metals (Stoner continuum). 
For materials with a collinear ground state it is possible to calculate the nearest neighbor exchange interaction of the quantum mechanical Heisenberg model using an energy mapping without spin-orbit effects \cite{Torelli2020}. In the general case however, there may not necessarily exist a meaningful energy mapping from a set of collinear magnetic configurations within DFT to the quantum mechanical Heisenberg model and one has to resort to a classical model of the magnetic interactions. 
Furthermore, any approach based on the Heisenberg model relies on the magnons being well described in terms of localized spins. This assumption becomes somewhat dubious for metallic systems, although it is often still possible to fit spin wave spectra to a Heisenberg model if enough exchange parameters are included in the fit \cite{Kudrnovsky2001}. For this reason, it is usually preferable to compute the Heisenberg exchange parameters directly from the magnetic force theorem (MFT) \cite{Liechtenstein1987,Bruno2003,Katsnelson2004} in the case of metals. Within the MFT framework, the Heisenberg exchange can be calculated in reciprocal space, allowing access to interactions on all length scales without any fundamental additional computational complexity.

In the adiabatic approach, it is assumed that the time dependent magnetization $\mathbf{m}(\mathbf{r}, t)$, which describes the magnon dynamics, can be treated as a classical variable on the time scale of the transverse magnetic fluctuations. This assumption implies that the underlying electronic degrees of freedom always should minimize the system energy given $\mathbf{m}(\mathbf{r}, t)$ on this time scale. Within such an approximation, the magnon dynamics are governed by a classical equation of motion (EOM) for $\mathbf{m}(\mathbf{r}, t)$, from which the magnon dispersion may be extracted \cite{Niu1998,Gebauer2000,Qian2002}. The material dependent parameters in the EOM can be computed based on a number of frozen magnon configurations generated within constrained DFT. However, it is not trivial to treat an arbitrary noncollinear magnetic structure consistently, and the frozen magnons have to be generated with care in order to reliably compute the magnon dispersion \cite{Bylander2000,Grotheer2001}. Although the adiabatic magnon EOM can be diagonalized in reciprocal space without any constraints on the spatial dependence of the magnetization \cite{Niu1998,Bylander2000}, it is often assumed that the magnetization can be represented in terms of localized (atomic) sites of volumes $V_i$ within which the direction of the magnetization is constant \cite{Halilov1998,Gebauer2000,Grotheer2001}. Within such a truncation of the spatial representation, the adiabatic approximation essentially reduces the magnon dynamics to those of the classical Heisenberg model \cite{Halilov1998,Niu1999}. Thus, when relying on the atomic sphere approximation or any similar spirited spatial partitioning of the magnetization, the adiabatic (frozen magnon) approach is essentially equivalent in nature to the energy mapping and MFT approaches, despite that the theoretical starting points and computational details may differ quite substantially. For this reason, all of the above approaches may be viewed as adiabatic, and they all share the characteristic that the magnon interaction with the Stoner continuum is neglected and that only the spin wave stiffness of itinerant ferromagnets can be treated exactly \cite{Katsnelson2000,Muniz2002}.

In order to appropriately capture the effects of the Stoner continuum on the magnon dispersion of itinerant ferromagnets, one needs either to map the problem onto a model accounting for the electronic degrees of freedom (e.g. a Hubbard model) or move on to the third approach, theoretical magnon spectroscopy. This approach relies on many-body perturbation theory (MBPT) or time-dependent density functional theory (TDDFT) to compute the full spectral function for the transverse magnetic excitations, which directly facilitates comparison and analysis of experimental results. In this work, we apply the linear response TDDFT (LR-TDDFT) method to compute the magnon dispersion of MnBi. Our methodology includes the renormalization effects of the Stoner continuum on the magnon dispersion, but does not aim to provide the reduced magnon lifetimes due to Landau damping as the lifetimes are above the current numerical resolution of our computational implementation. The framework of TDDFT in general \cite{Runge1984} and LR-TDDFT \cite{Gross1985} in particular have previously been successfully applied to extract magnon dispersion relations for several simple itinerant ferromagnets \cite{Savrasov1998,Buczek2011b,Lounis2011,Rousseau2012,Singh2018,Cao2017,Tancogne-Dejean2020,Skovhus2021}, just as it is the case for MBPT \cite{Aryasetiawan1999,Karlsson2000,SasIoglu2010,Okumura2019,Friedrich2020}. While the framework itself is formally exact, LR-TDDFT in practice relies on approximations to the exchange-correlation kernel, such as the Adiabatic Local Density Approximation (ALDA). The accuracy of the approximation is strongly dependent on the system at hand. For example, ALDA LR-TDDFT captures the magnon spectrum of bcc-Fe and hcp-Co rather accurately, whereas the bandwidth of the magnon spectrum of fcc-Ni is overestimated by a factor of two \cite{Buczek2011b,Rousseau2012,Singh2018,Skovhus2021}. The main reason for the inaccuracy is likely related to the inability of ALDA to correct for the Kohn-Sham exchange splitting, which is overestimated in LDA \cite{SasIoglu2010}.

In the context of ground state DFT, the disagreement between the Kohn-Sham spectrum and the quasi-particle spectrum may often be alleviated by the DFT+U approach \cite{Anisimov1991,Liechtenstein1995,Dudarev1998} where an on-site Hubbard repulsion is added, which tends to localize orbitals as well as increase band gaps and exchange splittings. In particular, Antropov et al. \cite{Antropov2014} have shown that a Hubbard correction is crucial in order to describe the ground state magnetic moments and the temperature-dependent magnetic anisotropy correctly in MnBi. Similarly, a Hubbard correction is also needed to reach an appropriate description of the structural properties of MnBi \cite{Shanavas2015}. However, in the context of LR-TDDFT it is not obvious how to include the Hubbard correction at the level of the exchange-correlation kernel, which introduces a mismatch between the kernel and the orbitals. This has severe consequences for the calculations since the delicate balance between the magnon spectrum and single-particle excitations is lost, leading to a gross violation of the Goldstone theorem in the combination of DFT+U and ALDA LR-TDDFT.

To remedy this violation, we apply a scalar rescaling of the exchange-correlation kernel, eliminating the mismatch between single-particle Stoner spectrum and magnon spectrum. The rescaling is fixed by the requirement of a gapless acoustic magnon mode and does not introduce any new free parameters apart from the ground state Hubbard correction. A scalar rescaling of the effective (screened) Coulomb interaction has also previously been adopted for theoretical magnon spectroscopy within MBPT \cite{SasIoglu2010,Okumura2019}, where it is prohibitively difficult to treat the effective interaction at full self-consistency with the single particle Stoner continuum \cite{Muller2016}. In this context, it has also been investigated whether a Hubbard correction (with rescaling of the effective interaction) could improve the faulty LDA magnon dispersion of fcc-Ni, with mixed success \cite{SasIoglu2010}. In the present work, it is shown that the experimentally measured magnon spectrum of MnBi is accurately captured within $\lambda$ALDA+U LR-TDDFT, whereas the neglect of Hubbard corrections leads to an overestimation of the optical magnon frequencies by at least a factor of two. In this way, it is demonstrated, that static correlation effects beyond the LDA are essential to include in order to correctly capture the inherent magnetic frustration in MnBi, and that a ground state Hubbard correction is a viable method in this regard. In particular, the correction strengthens the out-of-plane anti-ferromagnetic exchange interaction, and when artificially increasing the Hubbard correction beyond the experimental match, MnBi is imposed a phase transition to helical order. In addition, we demonstrate that MnBi undergoes a similar phase transition upon the introduction of either hole-doping (as one would expect in a Cr-alloy) or uniaxial compressive strain in the out-of-plane direction.
  
The paper is organized as follows. In Sec. \ref{sec:Theory} we outline the theoretical concepts underlying the LR-TDDFT framework for theoretical magnon spectroscopy. In Sec. \ref{sec:Methodology} we show how the Hubbard correction is included and how we determine the rescaling of the exchange-correlation kernel, which is required when combing LDA+U with ALDA LR-TDDFT. In addition, we also supply the computational details. In Sec. \ref{sec:results} we present the results of our study, including the effect of static correlations on ground and excited state properties of pristine MnBi as well as the effect of hole doping and uniaxial compressive strain on the magnon dynamics. Finally, in Sec. \ref{sec:discussion} we discuss how the results fit in as another piece in the puzzle of complex magnetic phases in the transition metal pnictide family and in Sec. \ref{sec:conclusion} we summarize our conclusions.

\section{Theory}\label{sec:Theory}

\subsection{The transverse magnetic susceptibility}

As exploited experimentally in inelastic neutron scattering and similar spectroscopic techniques, one may probe the fundamental excitations of a material by studying its response to external perturbations. In a nonrelativistic treatment, the linear order response in the transverse magnetic degrees of freedom can be fully characterized by a single response function, namely the transverse magnetic susceptibility. The susceptibility is given by the \textit{Kubo formula},
\begin{equation}
    \chi^{+-}(\mathbf{r}, \mathbf{r}', t-t') = - \frac{i}{\hbar} \theta(t-t') \langle\, [\hat{n}^{+}_0(\mathbf{r}, t), \hat{n}^{-}_0(\mathbf{r'}, t')] \,\rangle_0,
    \label{eq:trans mag susc kubo formula}
\end{equation}
where $\langle \cdot \rangle_0$ is the expectation value with respect to the ground state, $\theta(t-t')$ is the step function and $\hat{n}^{\pm}$ are the spin-raising and spin-lowering density operators respectively:
\begin{equation}
    \hat{n}^+(\mathbf{r}) = \hat{\psi}^{\dagger}_{\uparrow}(\mathbf{r}) \hat{\psi}_{\downarrow}(\mathbf{r}),
    \quad
    \hat{n}^-(\mathbf{r}) = \hat{\psi}^{\dagger}_{\downarrow}(\mathbf{r})         \hat{\psi}_{\uparrow}(\mathbf{r}).
\end{equation}
%
In Eq. \eqref{eq:trans mag susc kubo formula}, the operators carry the time-dependence of the interaction picture, $\hat n_0^\pm(\mathbf{r},t) \equiv e^{i\hat H_0t/\hbar} \, \hat n^\pm(\mathbf{r}) \, e^{-i\hat H_0t/\hbar}$, where $\hat H_0$ is the Hamiltonian of the unperturbed system, which in the case of MnBi has a ferromagnetic ground state. Furthermore, one may interchange the $+$ and $-$ indices in Eq. \eqref{eq:trans mag susc kubo formula} to define the susceptibility $\chi^{-+}$, but thanks to the relation $\chi^{-+}(\mathbf{r},\mathbf{r}',-\omega)=\chi^{+-*}(\mathbf{r},\mathbf{r}',\omega)$, it is sufficient to consider only $\chi^{+-}$. Beyond the nonrelativistic limit, one in general needs to consider the full four-component susceptibility tensor in order to characterize the magnetic modes of excitation \cite{Skovhus2021}. However, down to the experimental resolution of 6 meV, the available INS spectrum of MnBi does not exhibit any effects of the magnetic anisotropy \cite{Williams2016}, why such effects are neglected in the present study as well.

From the dissipative (anti-symmetric) part of the transverse magnetic susceptibility, one may extract the spectrum of induced excitations \cite{Skovhus2021}:
\begin{equation}
    S^{+-}(\mathbf{r}, \mathbf{r}', \omega) 
    = 
    - \frac{1}{2\pi i}\left[\chi^{+-}(\mathbf{r},\mathbf{r}',\omega) - \chi^{-+}(\mathbf{r}',\mathbf{r}, -\omega)\right].
    \label{eq:spectrum of induced excitations}
\end{equation}
%
For a ferromagnet (assumed spin-polarized along the $z$-direction), this spectrum determines the energy dissipation from weak perturbations in the transverse magnetic field $B_{\mathrm{ext}}^{x/y}(\mathbf{r}, t)$. Furthermore, $S^{+-}(\mathbf{r}, \mathbf{r}', \omega)$ constitutes a spectral function for the excited states which differ by a single unit of spin angular momentum compared to the ground state. In this way, one may use various spectroscopic techniques to extract $S^{+-}(\mathbf{r}, \mathbf{r}', \omega)$, permitting direct access to the transverse magnetic excitations of the system. Alternatively, $S^{+-}(\mathbf{r},  \mathbf{r}', \omega)$ can be computed by theoretical spectroscopy techniques, providing significant aid to the interpretation of measurements. More importantly, calculations of the transverse susceptibility allow one to rapidly scrutinize the effect of material modifications such as strain and doping, which may be time-consuming and costly to investigate experimentally. 

\subsection{Computing the transverse magnetic susceptibility within LR-TDDFT}

As a consequence of the Runge-Gross theorem \cite{Runge1984}, the time-dependent electronic structure of any material may be characterized in terms of an auxiliary Kohn-Sham system of non-interacting electrons where the electronic Coulomb repulsion is replaced by an effective (electromagnetic) potential. The susceptibility of the Kohn-Sham system can be evaluated using only quantities from a routine ground state DFT calculation \cite{HohenbergP.1973,Kohn1965,Barth1972,Rajagopal1973},
\begin{align}
    \chi^{+-}_{\mathrm{KS}}(\mathbf{r}, \mathbf{r}', \omega) = &\lim_{\eta \rightarrow 0^+} \frac{1}{N_k^2}\sum_{n\mathbf{k}} \sum_{m \mathbf{k}'} (f_{n\mathbf{k}\uparrow} - f_{m\mathbf{k}'\downarrow})
    \nonumber \\
    &\times \frac{\psi_{n\mathbf{k}\uparrow}^*(\mathbf{r}) \psi_{m\mathbf{k}'\downarrow}(\mathbf{r}) \,  \psi_{m\mathbf{k}'\downarrow}^*(\mathbf{r}') \psi_{n\mathbf{k}\uparrow}(\mathbf{r}')}{\hbar \omega - (\epsilon_{m\mathbf{k}'\downarrow}-\epsilon_{n\mathbf{k}\uparrow}) + i \hbar \eta},
    \label{eq:Kohn-Sham susc.}
\end{align}
where $N_k$ is the number of $k$-points, $\psi_{n\mathbf{k}s}(\mathbf{r})$ the Kohn-Sham orbital of band index $n$, $k$-point $\mathbf{k}$ and spin $s$, $f_{n\mathbf{k}s}$ the ground state occupancy and $\epsilon_{n\mathbf{k}s}$ the single-particle energy. In the adiabatic local density approximation (ALDA), the transverse magnetic susceptibility is directly related to the corresponding Kohn-Sham susceptibility through a single Dyson equation \cite{Gross1985},
\begin{align}
    \chi^{+-}(\mathbf{r}, \mathbf{r}', \omega) =& \chi^{+-}_{\mathrm{KS}}(\mathbf{r}, \mathbf{r}', \omega) + \int d\mathbf{r}_1 \, 
    \nonumber \\
    &\times \chi^{+-}_{\mathrm{KS}}(\mathbf{r}, \mathbf{r}_1, \omega) f^{-+}_{\mathrm{LDA}}(\mathbf{r}_1) \chi^{+-}(\mathbf{r}_1, \mathbf{r}', \omega),
    \label{eq:real-space Dyson eq.}
\end{align}
where $f^{-+}_{\mathrm{LDA}}(\mathbf{r}_1)=2 W_{\mathrm{xc,LDA}}^z(\mathbf{r})/n^z(\mathbf{r})$ is the transverse LDA kernel. Similar to the Kohn-Sham susceptibility, the kernel is given solely in terms of ground state quantities, namely the magnetic contribution to the LDA exchange-correlation potential, $W_{\mathrm{xc,LDA}}^z(\mathbf{r})$ (equal to the effective magnetic field up to a factor of $\mu_{\mathrm{B}}$), and the spin-polarization density $n^z(\mathbf{r})$. In this way, one can compute the many-body susceptibility of a given material within LR-TDDFT by computing $\chi^{+-}_{\mathrm{KS}}(\mathbf{r}, \mathbf{r}', \omega)$ and $f^{-+}_{\mathrm{LDA}}(\mathbf{r})$ for the LDA ground state and inverting the Dyson equation \eqref{eq:real-space Dyson eq.} within a suitable basis.

\subsection{Magnons and the plane wave susceptibility}

For periodic crystals (where $\hat{H}_0$ is invariant under lattice transformations $\hat{T}_{\mathbf{R}}$), one may characterize the transverse magnetic response in terms of the plane wave susceptibility,
\begin{equation}
    \chi^{+-}_{\mathbf{G}\mathbf{G}'}(\mathbf{q},\omega) 
        = 
        \iint \frac{d\mathbf{r} d\mathbf{r}'}{\Omega} e^{-i(\mathbf{G} + \mathbf{q}) \cdot \mathbf{r}} \chi^{+-}(\mathbf{r},\mathbf{r}',\omega) e^{i(\mathbf{G}' + \mathbf{q}) \cdot \mathbf{r}'},
\end{equation}
where $\Omega$ is the crystal volume, $\mathbf{G}$ and $\mathbf{G}'$ are reciprocal lattice vectors and $\mathbf{q}$ is a wave vector within the first Brillouin Zone (BZ). The plane wave susceptibility determines the coefficients for the linear order plane wave response $\propto e^{i([\mathbf{G}+\mathbf{q}]\cdot \mathbf{r} - \omega t)}$ to an external plane wave perturbation $\propto e^{i([\mathbf{G}'+\mathbf{q}]\cdot \mathbf{r} - \omega t)}$, a response which is diagonal in the reduced wave vector $\mathbf{q}$, thanks to the periodicity of the crystal. 

In full analogy with Eq. \eqref{eq:spectrum of induced excitations}, one can compute the plane wave spectrum of induced excitations \cite{Skovhus2021}:
\begin{equation}
    S^{+-}_{\mathbf{G}\mathbf{G}'}(\mathbf{q}, \omega) 
    = - \frac{1}{2 \pi i} \left[\chi^{+-}_{\mathbf{G}\mathbf{G}'}(\mathbf{q}, \omega) - \chi^{-+}_{-\mathbf{G}'-\mathbf{G}}(-\mathbf{q}, -\omega) \right].
\end{equation}
This spectrum can be decomposed into contributions from spin-lowering and spin-raising excitations encoded in the spectral functions $A^{+-}_{\mathbf{G}\mathbf{G}'}(\mathbf{q}, \omega)$ and $A^{-+}_{\mathbf{G}\mathbf{G}'}(\mathbf{q}, \omega)$ respectively:
\begin{equation}
    S^{+-}_{\mathbf{G}\mathbf{G}'}(\mathbf{q}, \omega) = A^{+-}_{\mathbf{G}\mathbf{G}'}(\mathbf{q}, \omega) - A^{-+}_{-\mathbf{G}'-\mathbf{G}}(-\mathbf{q}, -\omega).
\end{equation}
Both of these spectral functions have peaks at frequencies $\hbar \omega = E_{\alpha} - E_{0}$ corresponding to the transition energies between the excited states $|\alpha\rangle$ and the ground state $|\alpha_0\rangle$. The contribution from each excited state is weighted by the reciprocal space pair densities $n^{j}_{\alpha\alpha'}(\mathbf{G}+\mathbf{q})$,
\begin{align}
    A^{jk}_{\mathbf{G}\mathbf{G}'}(\mathbf{q}, \omega) 
    =& \frac{1}{\Omega}\sum_{\alpha \neq \alpha_0} n^j_{0\alpha}(\mathbf{G}+\mathbf{q}) n^k_{\alpha0}(-\mathbf{G}'-\mathbf{q})
    \nonumber \\
    &\times \delta\big(\hbar \omega - (E_\alpha - E_0)\big),
\end{align}
where $\delta(\hbar \omega - \Delta E)$ denotes the Dirac delta-function. The reciprocal space pair densities are Fourier transforms of the real-space pair densities $n^j_{\alpha\alpha'}(\mathbf{r})=\langle \alpha|\hat n^j(\mathbf{r})|\alpha'\rangle$. As the effect of the spin-lowering and spin-raising operators is to remove a spin-up or a spin-down electron at the position $\mathbf{r}$ respectively and replace it with an electron of the opposite spin, the $A^{+-}$ and $A^{-+}$ spectral functions include only excited states for which the spin angular momentum has been either lowered or raised by a single unit compared to the ground state. Furthermore, the reciprocal space pair densities are only non-zero for excited states which differ in crystal momentum by $\hbar \mathbf{q}$ compared to the ground state. In this way, $A^{+-}_{\mathbf{G}\mathbf{G}'}(\mathbf{q}, \omega)$ and $A^{-+}_{\mathbf{G}\mathbf{G}'}(\mathbf{q}, \omega)$ encode quasi-particle excitations of energy $\hbar \omega$ and crystal momentum $\hbar \mathbf{q}$, excitations that carry a single unit of spin angular momentum.
In turn, the plane wave spectrum $S^{+-}_{\mathbf{G}\mathbf{G}'}(\mathbf{q}, \omega)$ of a ferromagnet assumed spin-polarized in the $z$-direction encodes majority-to-minority magnons at positive frequencies and minority-to-majority magnons at negative frequencies. Concerning the reciprocal lattice vectors, it is useful to focus on the diagonal $S^{+-}_{\mathbf{G}}(\mathbf{q}, \omega) = S^{+-}_{\mathbf{G}\mathbf{G}}(\mathbf{q}, \omega)$. As the reciprocal lattice vector $\mathbf{G}$ represents the local field component of the spin-flipping pair densities, it also represents the local spin texture of the excitations within the unit cell. Accordingly, the acoustic magnon mode will appear in the spectrum $S^{+-}_{\mathbf{G}}(\mathbf{q}, \omega)$ for values of $\mathbf{G}$ for which the magnetic atoms of the unit cell are in phase. Likewise, the optical mode of MnBi will be visible in $S^{+-}_{\mathbf{G}}(\mathbf{q}, \omega)$ for values of $\mathbf{G}$, for which the two Mn atoms in the unit cell are out of phase.

\section{Methodology}\label{sec:Methodology}

\subsection{Ground state Hubbard correction}

Although one would not normally expect strong correlation effects in metals, previous \textit{ab initio} studies of MnBi have shown that the LDA provides an insufficient description of properties ranging from lattice constants and equilibrium magnetic moments to the temperature dependent magnetic anisotropy and magneto-optical effects \cite{Antropov2014,Shanavas2015}. For these properties, a significant improvement is obtained when including a Hubbard-like on-site correction to the 3\textit{d} electronic orbitals of Mn. In this study, we apply the rotationally invariant Dudarev LDA+U scheme \cite{Dudarev1998}, where the Coulomb interaction amongst the 3\textit{d} electrons is corrected with a single effective Hubbard parameter $U_{\mathrm{eff}}=U-J$. The Hubbard correction favors idempotency and splits the localized majority and minority 3\textit{d} bands by a value similar to $U_{\mathrm{eff}}$.

\subsection{Hubbard correction and LR-TDDFT}

In order to relate the Kohn-Sham susceptibility to the many-body susceptibility by means of the Dyson equation \eqref{eq:real-space Dyson eq.}, the xc-kernel needs to be derived from a time-dependent xc-potential that reproduces the correct ground state spin densities and effective potentials in the static limit (in the absence of an external perturbation). This implies, that the ALDA xc-kernel cannot be used to calculate the transverse magnetic susceptibility based on a LDA+U ground state calculation.

\subsubsection{Collective enhancement}
To elaborate on how the Dyson equation \eqref{eq:real-space Dyson eq.} breaks down in practice, we invert it in the plane wave basis,
\begin{equation}
    \chi^{+-}_{[\mathbf{G}]}(\mathbf{q}, \omega) =\left(1 - \chi^{+-}_{\mathrm{KS}}(\mathbf{q}, \omega) f^{-+}_{\mathrm{LDA}} \right)^{-1}_{[\mathbf{G}]} \chi^{+-}_{\mathrm{KS}, [\mathbf{G}]}(\mathbf{q}, \omega),
    \label{eq:reciprocal space dyson}
\end{equation}
and compare it to the Dyson equation for a homogeneous electron gas (HEG), for which the susceptibility in reciprocal space is a scalar function of the wave number $q$ \cite{Niesert2011,Friedrich2020}:
\begin{equation}
    \chi^{+-}(q, \omega) =\frac{\chi^{+-}_{\mathrm{KS}}(q, \omega)}{1 - \chi^{+-}_{\mathrm{KS}}(q, \omega) f^{-+}_{\mathrm{LDA}}}.
    \label{eq:HEG Dyson eq}
\end{equation}
For $\chi^{+-}(q, \omega)$, $\chi^{+-}_{\mathrm{KS}}(q, \omega)$ and the diagonal elements of $\chi^{+-}_{\mathbf{G}\mathbf{G}'}(\mathbf{q},\omega)$, the spectrum of induced excitations is proportional to the imaginary part of the susceptibility \cite{Skovhus2021}: $S^{+-}_{\mathbf{G}}(\mathbf{q}, \omega) = - \mathrm{Im}\big[ \chi^{+-}_{\mathbf{G}\mathbf{G}}(\mathbf{q}, \omega) \big] / \pi$. The Kohn-Sham spectrum $S^{+-}_{\mathrm{KS}}(q, \omega)$ forms a continuum of Stoner pair excitations, that is, single-particle electron-hole pairs involving a single spin-flip. Through Eq. \eqref{eq:HEG Dyson eq}, the Stoner excitations are carried over to the many-body spectrum $S^{+-}(q, \omega)$ at a renormalized spectral intensity determined by the real part of the denominator. Apart from the Stoner excitations, Eq. \eqref{eq:HEG Dyson eq} also permits a new type of excitation at the roots of the real part of the denominator. Inspired by this fact, we introduce the inverse enhancement function (IEF):
\begin{equation}
    \kappa^{+-}(q, \omega) \equiv \mathrm{Re}\left[ \frac{\chi^{+-}_{\mathrm{KS}}(q, \omega)}{\chi^{+-}(q, \omega)} \right] = 1 - \mathrm{Re}\left[\chi^{+-}_{\mathrm{KS}}(q, \omega) \right] f^{-+}_{\mathrm{LDA}}.
    \label{eq:HEG inverse enhancement function}
\end{equation}
This function determines the collective enhancement of the many-body spectrum due to the electron-electron interaction described by $f^{-+}_{\mathrm{LDA}}$. Outside the Stoner continuum, where $\mathrm{Im}\big[\chi^{+-}_{\mathrm{KS}}(q, \omega)\big]=0$, collective magnon excitations appear at roots of the IEF. Furthermore, if the IEF vanishes (or nearly vanishes) inside the Stoner continuum, the Stoner pair excitations in the vicinity are enhanced. Whereas the Stoner continuum for the spin-polarized HEG is gapped at $q=0$ by the exchange splitting energy $\Delta_{\mathrm{x}}$, the many-body spectrum exhibits a so-called Goldstone magnon mode with $\omega_{q=0}=0$. As the IEF defines the magnon dispersion exactly outside the Stoner continuum, the Goldstone theorem thus dictates that $\kappa^{+-}(0, 0) = 0$ for the spin-polarized HEG. In a similar spirit, we may introduce the following expression as the inverse enhancement function for periodic crystals:
\begin{equation}
    \kappa^{+-}_{\mathbf{G}}(\mathbf{q}, \omega) \equiv \mathrm{Re}\left[ \frac{\chi^{+-}_{\mathrm{KS},\mathbf{G}\mathbf{G}}(\mathbf{q}, \omega)}{\chi^{+-}_{\mathbf{G}\mathbf{G}}(\mathbf{q}, \omega)} \right].
    \label{eq:inverse enhancement function}
\end{equation}
Whereas the IEF for the HEG \textit{determines} the collective enhancement in the Dyson equation \eqref{eq:HEG Dyson eq}, $\kappa^{+-}_{\mathbf{G}}(\mathbf{q}, \omega)$ provides a post hoc \textit{characterization} of the enhancement. Nevertheless, we find that the magnon condition $\kappa^{+-}_{\mathbf{G}}(\mathbf{q}, \omega)=0$ provides a good approximation for the peak position of magnon excitations away from dense parts of the Stoner continuum for MnBi as well as for the simpler itinerant ferromagnets Fe, Ni and Co. More generally, it is the matrix $\left(1 - \chi^{+-}_{\mathrm{KS}}(\mathbf{q}, \omega) f^{-+}_{\mathrm{LDA}} \right)^{-1}_{[\mathbf{G}]}$ that determines the enhancement in the Dyson equation \eqref{eq:reciprocal space dyson}. Thus, it is somewhat surprising that the collective enhancement can be effectively characterized using a scalar function for each local field component. We attribute this success to the fact that the definition in Eq. \eqref{eq:inverse enhancement function} involves the fully enhanced $\chi^{+-}_{\mathbf{G}\mathbf{G}}(\mathbf{q}, \omega)$, thus implicitly accounting for the contribution of all plane wave components.

\subsubsection{Gap error and rescaling of the ALDA kernel}

\begin{figure*}[ht]
    \centering
    \includegraphics{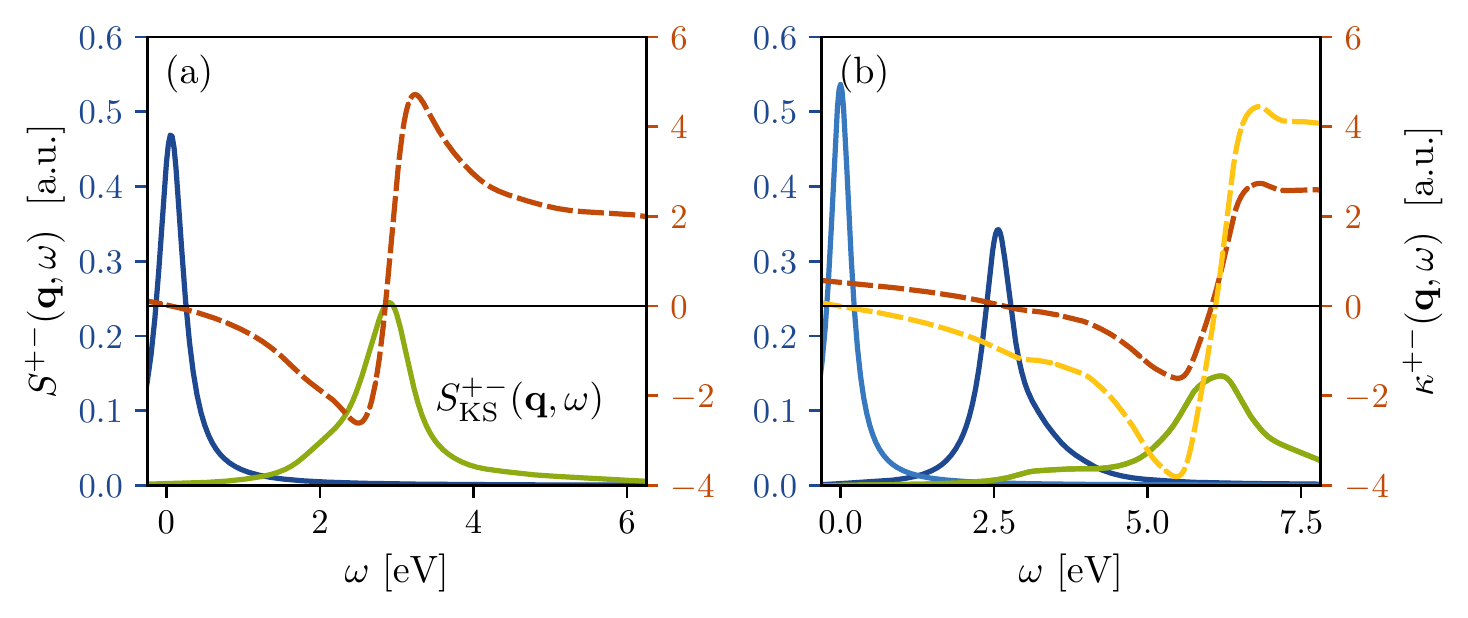}
    \caption{Collective enhancement of the Goldstone mode in MnBi at $\mathbf{q}=\mathbf{0}$. The macroscopic Stoner spectrum $S^{+-}_{\mathrm{KS}}(\mathbf{q}, \omega)=S^{+-}_{\mathrm{KS},\mathbf{G}=\mathbf{0}}(\mathbf{q}, \omega)$ is shown in green, the inverse enhancement function $\kappa^{+-}_{\mathbf{G}=\mathbf{0}}(\mathbf{q}, \omega)$ in red and the macroscopic many-body spectrum $S^{+-}_{\mathbf{G}=\mathbf{0}}(\mathbf{q}, \omega)$ in blue.
    (a) ALDA collective enhancement.
    (b) ALDA+U collective enhancement. Red and dark blue show the enhancement using the bare ALDA kernel calculated for the LDA+U ground state using $U_{\mathrm{eff}}=3$ eV. The yellow and light blue colors shows the enhancement using the $\lambda$ALDA+U kernel scaled to fulfill the Goldstone condition $\kappa^{+-}_{\mathbf{0}}(\mathbf{0}, 0)=0$.
    }
    \label{fig:collective_enhancement}
\end{figure*}
In Fig. \ref{fig:collective_enhancement}(a) the collective enhancement is illustrated for the Goldstone mode of MnBi at $\mathbf{q}=\mathbf{0}$, calculated using the ALDA kernel on the LDA ground state. The Kohn-Sham spectrum has a well-defined peak at 2.9 eV, corresponding to the LDA exchange splitting. At the Kohn-Sham peak position, the IEF exhibits a pole-like feature with a shape that closely resembles $\mathrm{Re}\left[ \chi^{+-}_{\mathrm{KS},\mathbf{G}\mathbf{G}}(\mathbf{q}, \omega) \right]$, which in turn forms a Kramers-Kronig pair with the Kohn-Sham spectrum. In this way, the plane wave IEF behaves exactly as one would expect from the HEG, where $\mathrm{Re}\left[ \chi^{+-}_{\mathrm{KS}}(q, \omega) \right]$ carries all the frequency dependence in $\kappa^{+-}(q, \omega)$, as seen from Eq. \eqref{eq:HEG inverse enhancement function}. 
Far away from the Stoner continuum, the Kohn-Sham susceptibility vanishes and the IEF goes to unity. Because the IEF takes negative values for frequencies just below the exchange splitting energy, it also obtains a root below the Kohn-Sham peak, which in the macroscopic case ($\mathbf{G}=\mathbf{0}$) gives rise to the acoustic (Goldstone) magnon mode. For the ALDA kernel applied to the the LDA ground state shown in Fig. \ref{fig:collective_enhancement}(a), the acoustic magnon frequency goes to zero in the long-range limit, thus fulfilling the Goldstone theorem (at least within some tenths of meV using the applied computational parameters).

When applying the ALDA kernel to the LDA+U ground state however, the Goldstone theorem is violated. As shown in Fig. \ref{fig:collective_enhancement}(b), including a Hubbard correction of $U_{\mathrm{eff}}=3$ eV increases the Kohn-Sham peak position to 6.2 eV, i.e. the effective exchange splitting increases by a value $\simeq U_{\mathrm{eff}}$. However, the ALDA kernel does not change sufficiently to accommodate the increased exchange splitting, resulting in a gap error of $\omega_{\Gamma}=2.6$ eV. In the HEG limit, the ALDA kernel amounts to a scalar entity giving the effective interaction strength amongst the single-particle Stoner excitations. Furthermore, the HEG magnon spectrum consist of a single mode, i.e.  only a unique root in $\kappa^{+-}(q, \omega)$ corresponds to a collective excitation. Consequently, the Goldstone condition $\kappa^{+-}(0, 0) = 0$ fixes the scalar ALDA kernel $f^{-+}_{\mathrm{LDA}}$ in Eq. \eqref{eq:HEG Dyson eq}. Conversely, if the Goldstone condition is violated, it is consistent with the ALDA to simply scale the value of $f^{-+}_{\mathrm{LDA}}$ to satisfy $\kappa^{+-}(0, 0) = 0$. In a similar spirit, we rescale the plane wave kernel $f^{-+}_{\mathrm{LDA},[\mathbf{G}]} \rightarrow \lambda f^{-+}_{\mathrm{LDA},[\mathbf{G}]}$ such as to satisfy the approximate Goldstone condition $\kappa^{+-}_{\mathbf{G}=\mathbf{0}}(\mathbf{q}=\mathbf{0}, \omega=0)=0$ whenever a Hubbard correction has been applied. Henceforward, we will refer to the kernel calculated in this way as the $\lambda$ALDA+U kernel. As seen in Fig. \ref{fig:collective_enhancement}(b), this rescaling increases the intensity of the pole-like feature in $\kappa^{+-}_{\mathbf{G}}(\mathbf{q}, \omega)$ stemming from the Stoner continuum and thus effectively moves the root as well as the magnon peak position to zero frequency. Since the Hubbard correction mainly affects the 3$d$ electrons, it is a somewhat naive approach to rescale the entire ALDA kernel, but as we will show in Sec. \ref{sec:results - dynamic correlation effects}, the inclusion of a Hubbard correction in the $\lambda$ALDA+U approach leads to significant improvements of the magnon dispersion when compared to experiment. As a result, it seems that the $\lambda$ALDA+U approach can be a valuable tool for including correlation effects within LR-TDDFT in a simple and transparent way. In the literature, more advanced approaches have been developed to reclaim the self-consistency between the exchange-correlation kernel and Kohn-Sham susceptibility in order to strictly satisfy the Goldstone theorem, also when it is violated due to numerical limitations \cite{Buczek2011b,Lounis2011,Rousseau2012}. For a broader application of the LR-TDDFT methodology within e.g. the transition metal pnictide family, it would be of high interest to make a comparative study of different strategies to satisfy the Goldstone condition in calculations with Hubbard corrections in the ground state.

Analogously to the $\lambda$ALDA+U approach, a global rescaling of the effective interaction has also previously been applied to satisfy the Goldstone theorem in MBPT calculations of the magnon dispersion \cite{SasIoglu2010,Okumura2019}. Based on the $GW$ approximation for the self-energy and a static limit of the random phase approximation for the screened Coulomb potential $W$, on may rescale $W\rightarrow\lambda W$ to satisfy the Goldstone theorem when using the LDA+U Green's function $G_0$. This approach leads to an improved magnon stiffness for fcc-Ni as compared to a LDA based calculation \cite{SasIoglu2010}. However, for fcc-Ni the exchange splitting is overestimated already in the LDA and inclusion of the Hubbard correction worsens the overall magnon bandwidth (and thus the short-range transverse spin correlations), contrary to the present case for MnBi. In MBPT, the justification for using a global rescaling of the effective interaction rests on similar grounds as in the $\lambda$ALDA+U approach, see e.g. the comparison to the one-band Hubbard model presented in Ref. \cite{Muller2016}. There is an important distinction however: Even when using the LDA Kohn-Sham susceptibility as a starting point, a gap error is obtained in MBPT, due to the inconsistency with the $GW$ self-energy approximation \cite{Muller2016}.

\subsection{Computational details}

All calculations in this study have been carried out using the open-souce GPAW code \cite{Mortensen2005,Enkovaara2010}, with which the plane wave susceptibility $\chi^{+-}_{\mathbf{G}\mathbf{G}'}(\mathbf{q},\omega)$ can be computed within LR-TDDFT \cite{Skovhus2021}. In principle, the PAW implementation yields all-electron accuracy, but in practise it is not possible to construct a complete basis of projector functions. In the applied PAW-setups, only the Mn 4$s$ and 3$d$ as well as the Bi 6$s$, 6$p$ and 5$d$ orbitals are included as valence states in the band summation of Eq. \eqref{eq:Kohn-Sham susc.}. In addition to the valence states, 8 empty shell bands per atom are included as well. Furthermore, a (30, 30, 18) $\Gamma$-centered Monkhorst-Pack (MP) grid is used along with a plane wave cutoff of 750 eV for the plane wave basis in Eq. \eqref{eq:reciprocal space dyson}. These parameters were chosen based of an extensive convergence study of the magnon dispersion in the itinerant ferromagnets Fe, Ni and Co \cite{Skovhus2021} and we have checked that these parameters lead to well converged results for MnBi as well.

Calculations have been carried out in the NiAs-type crystal structure, using the experimental room temperature lattice constants ($a=4.287\,\mathrm{\AA}$, $c=6.117\,\mathrm{\AA}$) for LR-TDDFT calculations. It should be noted that the reference inelastic neutron scattering data available is taken at 5 K \cite{Williams2016} and that the MnBi structure is contracting when cooling, all the way from 293 K, through the structural phase transition at 90 K, down to at least 20 K \cite{McGuire2014}. Nevertheless, we expect a fair comparison between theory and experiment, when using the room temperature crystal structure for the simulations. Occasional comparisons of magnon dispersion relations with calculations performed in the orthorhombic crystal structure reported at 80 K \cite{McGuire2014} 
have been carried out, without any noteworthy qualitative differences as a result.

In order to reliably extract the magnon dispersion, the Kohn-Sham continuum, which is sampled on a finite $k$-point mesh, needs to be broadened into a continuum. In practise this is done by leaving $\eta$ as a small, but finite parameter in Eq. \eqref{eq:Kohn-Sham susc.}. Through the average frequency displacement technique described in \cite{Skovhus2021}, it was found that the low frequency Kohn-Sham spectrum of MnBi is effectively broadened into a continuum for values of $\eta \geq 100$ meV using the chosen $k$-point grid. Unfortunately, the broadening parameter $\eta$ not only dictates the minimum width of features in the Kohn-Sham spectrum, but in the many-body spectrum as well. Using $\eta=100$ meV, the spectral width of the magnons will be of the same order of magnitude as the MnBi magnon bandwidth. 
\begin{figure*}[ht]
    \centering
    \includegraphics{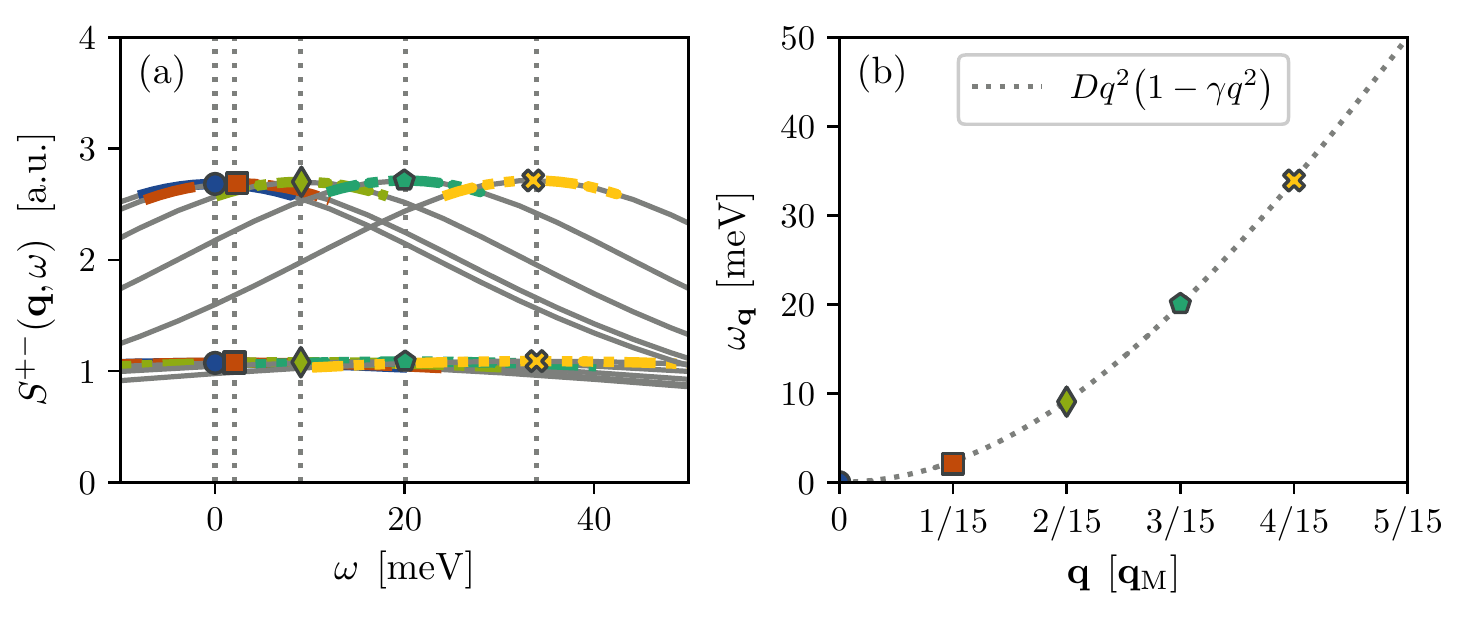}
    \caption{Extraction of the magnon stiffness from the $\lambda$ALDA+U transverse magnetic excitation spectrum with $U_{\mathrm{eff}}=3\,\mathrm{eV}$. 
    (a) Magnon spectrum along the $\Gamma\rightarrow\mathrm{M}$ path calculated on a $(30, 30, 18)$ $\Gamma$-centered MP-grid using $\eta=100\,\mathrm{meV}$ (lower set of lineshapes) and a $(60, 60, 36)$ $\Gamma$-centered MP-grid using $\eta=40\,\mathrm{meV}$ (upper set of lineshapes). Magnon energies (scatter points) are identified as peak positions from a quadratic fit to the maximum of each Lorentzian lineshape (colored lines). Vertical dotted lines indicate the magnon energies from the sparse $k$-point grid for visual comparison.
    (b) Extracted magnon dispersion from the calculation using the sparse $k$-point grid along with a biquadratic fit, $\omega_{\mathbf{q}}=D q^2\big(1-\gamma q^2\big)$, to the dispersion.}
    \label{fig:spin_stiffness_example}
\end{figure*}
Consequently, it becomes difficult for the human eye to discern the magnon dispersion directly from the spectrum, as illustrated in Fig. \ref{fig:spin_stiffness_example}(a), and it is not possible to extract a physical lineshape for the magnons. This means that the Landau damping of the MnBi magnon modes cannot be studied with the present numerical resolution in GPAW. To elucidate such effects in MnBi, additional methodology would need to be implemented, such as the tetrahedron integration or analytic continuation techniques, which both previously have been proved effective for \textit{ab initio} treatments of itinerant ferromagnets \cite{Buczek2011b,Friedrich2020}.  However, the available INS spectra do not exhibit any clear itinerant electron effects within the instrument resolution \cite{Williams2016} and the study of such effects in MnBi is left for future work. Using the present methodology, we instead focus on the LR-TDDFT magnon dispersion, which includes magnon renormalization effects from the Stoner continuum in contrast to a Heisenberg model approach. To extract magnon energies, we sample the spectrum on a frequency grid with a spacing $\delta \omega \leq \eta / 8$ and compute the magnon frequency as the maximum in a parabolic fit to the magnon spectral peak. In Fig. \ref{fig:spin_stiffness_example}(a), the magnon peak positions extracted from the (30, 30, 18) $k$-point grid are compared to similarly extracted peak positions on a (60, 60, 36) grid, where it was possible to reduce $\eta$ to a value of 40 meV without compromising the Kohn-Sham continuum. Clearly, it is not necessary to go beyond the (30, 30, 18) $k$-point grid in order to converge the magnon dispersion and a value of $\eta=100$ meV will be used throughout the remainder of the paper. 

As a final note, we extract the magnon stiffness $D$ along a given direction in reciprocal space by performing a biquadratic fit to the magnon peak positions of the 4 shortest $\mathbf{q}$-vectors along the path in question (excluding the $\Gamma$-point). This procedure is illustrated in Fig \ref{fig:spin_stiffness_example}(b) for the $\Gamma\rightarrow\textrm{M}$ direction.

\section{Results}\label{sec:results}

\subsection{Correlation effects in the ground state}\label{sec:ground state results}
\begin{figure*}[ht]
    \centering
    \includegraphics{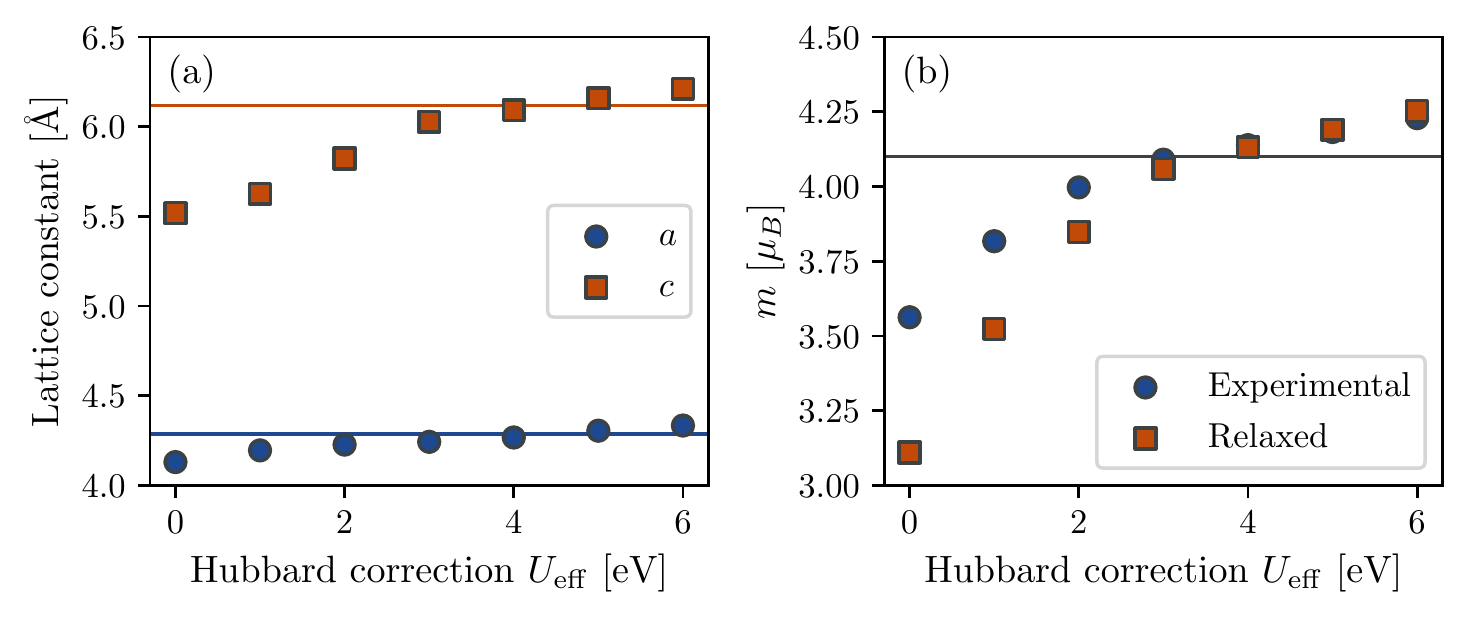}
    \caption{{ Effects of Hubbard corrections on lattice parameters and magnetization. 
    (a) LDA+U lattice constants as a function of $U_{\mathrm{eff}}$, calculated using the Atomic Simulation Recipes (ASR) \cite{Gjerding2021}. Solid lines indicate the experimental lattice constants at room temperature \cite{McGuire2014}.
    (b) Magnetization per Mn atom as a function of $U_{\mathrm{eff}}$ using the experimental and LDA+U crystal structures respectively. The experimental magnetization at 5 K \cite{McGuire2014} is shown as a reference.
    }}
    \label{fig:ground_state_properties}
\end{figure*}
As argued in Refs. \cite{Antropov2014,Shanavas2015}, it is necessary to include a Hubbard correction in the DFT ground state description of MnBi in order to capture both structural and magnetic properties. In Fig. \ref{fig:ground_state_properties}(a) we compare the lattice constants in the NiAs-type crystal structure calculated with LDA(+U) to the experimental room temperature crystal structure. Without a Hubbard correction, we find the lattice constants $a=4.131\,\mathrm{\AA}$ and $c=5.519\,\mathrm{\AA}$, which are both underestimated compared to the experimental lattice constants at room temperature, differing by $\Delta a=0.156\,\mathrm{\AA}$ and $\Delta c=0.598\,\mathrm{\AA}$ respectively. The temperature effect only provides a minor contribution to this difference. Experimentally, $a$ and $c$ are found to contract $<0.02\,\mathrm{\AA}$ and $<0.06\,\mathrm{\AA}$ respectively when the material is cooled from room temperature to 20 K \cite{McGuire2014}. Thus, the error mainly resides within the exchange-correlation functional, which in the case of LDA significantly underestimates the lattice constants, especially the out-of-plane lattice constant $c$, which determines the nearest neighbour distance between the Mn atoms. In the case of LDA+U however, reasonable lattice constants are obtained for values of $U_{\mathrm{eff}}$ within the range of $3-5$ eV, with $U_{\mathrm{eff}}=4$ eV providing the best match to experiment.
In Fig. \ref{fig:ground_state_properties}(b), we show the LDA(+U) ground state magnetization per Mn atom
calculated for the experimental and relaxed DFT crystal structures respectively. In LDA, the magnetization is significantly underestimated with a value of $3.56\,\mu_{\mathrm{B}}$ (to be compared with the experimental value of $4.1\,\mu_{\mathrm{B}}$ \cite{McGuire2014}). If we instead use the LDA lattice constants, the situation is even worse as the magnetization takes a value of $3.11\,\mu_{\mathrm{B}}$. Again, this deficiency is amended upon the inclusion of a Hubbard correction and a reasonable agreement with experiment is found for values of $U_{\mathrm{eff}}$ in the range of 2-5 eV. As an alternative to Hubbard corrections, one can also obtain consistent improvements of ground state properties using a GGA functional \cite{Shanavas2015}. However, whereas the GGA in-plane lattice constant matches experiments well, Hubbard corrections are still needed in order to capture the out-of-plane lattice constant and the magnetization.

\begin{figure*}[ht]
    \centering
    \includegraphics{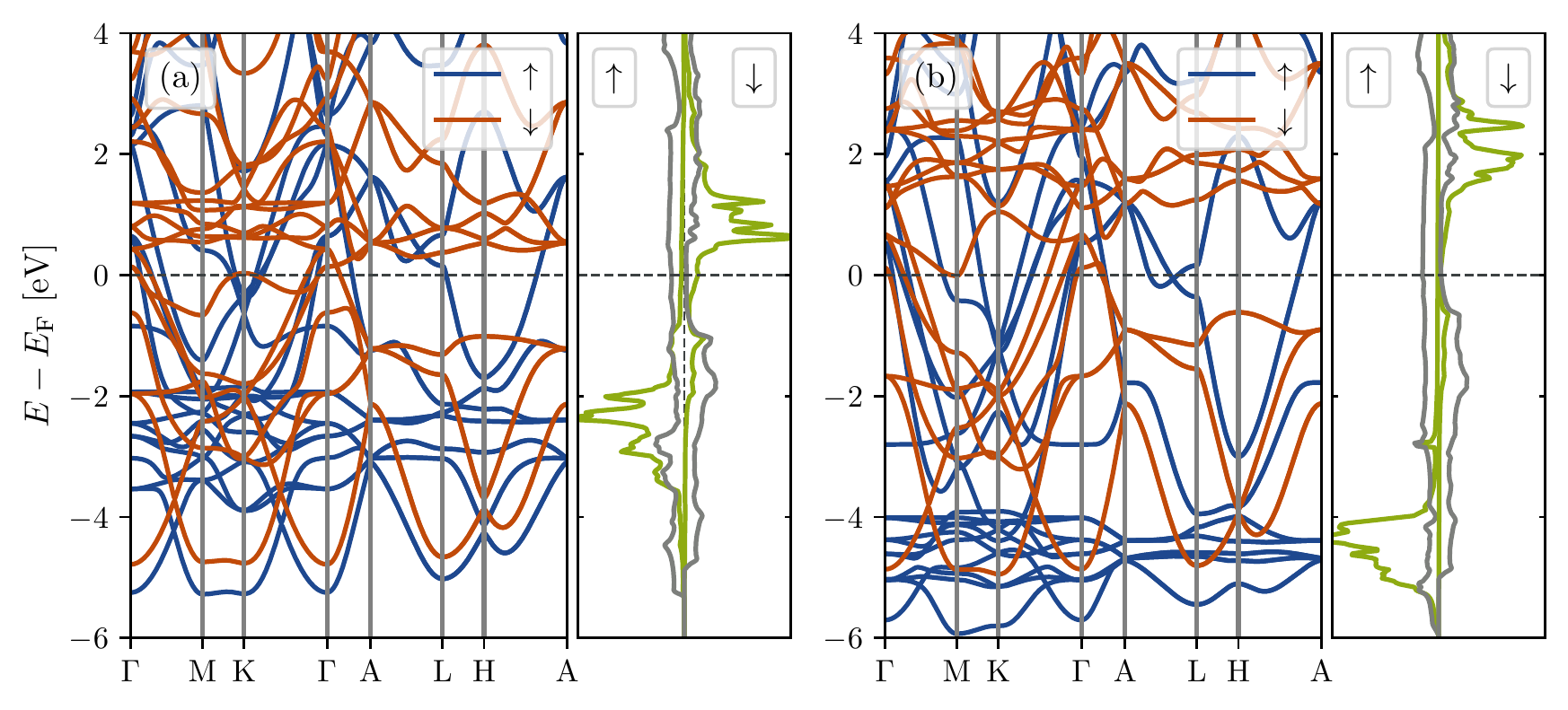}
    \caption{Kohn-Sham electronic band structure and projected density of states (PDOS). Projections in green indicate the Mn 3$d$ orbitals, whereas the grey projections show the sum of contributions from the Mn 4$s$, Mn 4$p$, Bi 6$s$ and Bi 6$p$ orbitals.
    (a) LDA band structure and PDOS. 
    (b) LDA+U band structure and PDOS ($U_{\mathrm{eff}}=3$ eV).}
    \label{fig:band_structures_with_pdos}
\end{figure*}
Similar to the lattice constant $c$ in Fig. \ref{fig:ground_state_properties}(a), the magnetization in Fig. \ref{fig:ground_state_properties}(b) increases monotonically with $U_{\mathrm{eff}}$, but most rapidly so for values of $U_{\mathrm{eff}}$ up to 3 eV. This can be understood based on the band structures and projected density of states shown in Fig. \ref{fig:band_structures_with_pdos}. Generally speaking, the Hubbard correction increases the exchange splitting, moving the minority bands up in energy and the majority bands down in energy compared to the Fermi level, hence the monotonic increase in magnetization. For the Kohn-Sham band structure in the vicinity of the Fermi level, one may separate the bands into two groups based on the orbital projections. There is a group of narrow bands consisting mainly of Mn 3$d$ orbitals as well as a group of more dispersive bands of mixed orbital character. Due to the nature of the Hubbard correction, it mostly affects the narrow Mn 3$d$ bands. For $U_{\mathrm{eff}}=3$ eV, the Mn 3$d$ minority bands are moved $\sim 1$ eV up and the Mn 3$d$ majority bands are moved $\sim 2$ eV down in energy with respect to the Fermi level. That is, the Mn 3$d$ exchange splitting is increased by a value $\sim U_{\mathrm{eff}}$. For the Mn 3$d$ majority bands, the energy shift is not associated with a change in the occupancies, as the bands are placed 2-4 eV below the Fermi level  in the LDA band structure. In contrast, there are Mn 3$d$ minority states situated around the Fermi level of the LDA band structure, some of them partially occupied. Since the Hubbard correction increases the exchange splitting, the Mn 3$d$ minority bands are shifted away from the Fermi level and become fully unoccupied at $U_{\mathrm{eff}}=3$ eV. Thus, for values of $U_{\mathrm{eff}}$ above 3 eV, the increasing magnetization is determined solely from the more dispersive bands of mixed orbital character, which are less affected by the correction. As a consequence, the magnetization increases only slowly with $U_{\mathrm{eff}}$ above 3 eV.

\subsection{Correlation effects in the magnon dynamics}\label{sec:results - dynamic correlation effects}

\begin{figure*}[ht]
    \centering
    \includegraphics{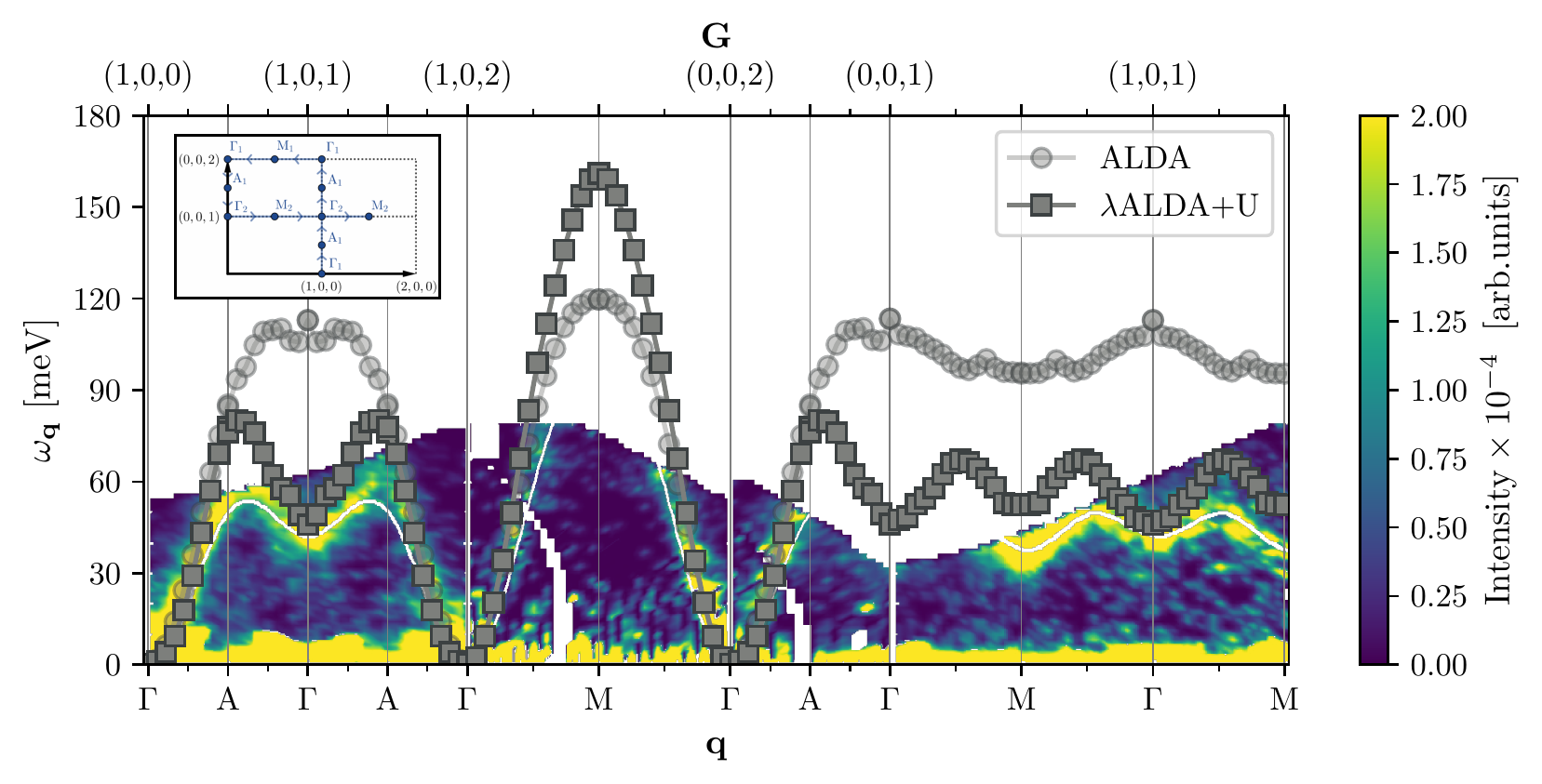}
    \caption{Theoretical ALDA and $\lambda$ALDA+U ($U_{\mathrm{eff}}=3\,\mathrm{eV}$) magnon dispersion relations plotted on top of a heat map of the transverse magnetic structure factor as probed by inelastic neutron scattering \cite{Williams2016}. In the inset, the wave vector path, $\mathbf{G}+\mathbf{q}$, is shown, given in coordinates of the reciprocal lattice vectors of the hexagonal NiAs crystal structure.}
    \label{fig:full_magnon_dispersion_expSpec}
\end{figure*}
While the effect of Hubbard corrections on the ground state properties have been discussed in Refs. \cite{Antropov2014,Shanavas2015}, we extend the discussion to include also the influence on the dynamic susceptibility of MnBi. As shown above, the LDA Kohn-Sham spectrum in Fig. \ref{fig:collective_enhancement}(a) has a well-defined Stoner peak at 2.9 eV. From the band structure and projected density of states shown in Fig. \ref{fig:band_structures_with_pdos}(a), we see that the origin of this peak is the narrow Mn 3$d$ bands, which dominate the spin dynamics of the system as expected. However, as reflected in Fig. \ref{fig:band_structures_with_pdos}, the Fermi surface includes contributions from the dispersive bands of mixed orbital character and these will have an important influence on the fundamental spin excitations as well. As the exchange splitting between the Mn 3$d$ bands grows, a non-negligible contribution from the dispersive bands becomes visible below the main Stoner peak, see Fig. \ref{fig:collective_enhancement}(b). For values of $U_{\mathrm{eff}}$ above 3 eV, it would be tempting to apply a localized (half-)integer spin model based on the gapped Mn 3$d$ bands, but this would largely neglect the itinerant electronic effects introduced by the dispersive bands crossing the Fermi level. For example, the magnon dispersion is expected to exhibit  Landau damping, which originates from the coupling of the collective Mn $d$-band excitations to the Stoner continuum originating from states in the vicinity of the Fermi level. In contrast, LR-TDDFT includes contributions from all bands on the same footing and is expected to capture both the Landau damping as well as the associated renormalization of the magnon dispersion.

\begin{figure*}[ht]
    \centering
    \includegraphics{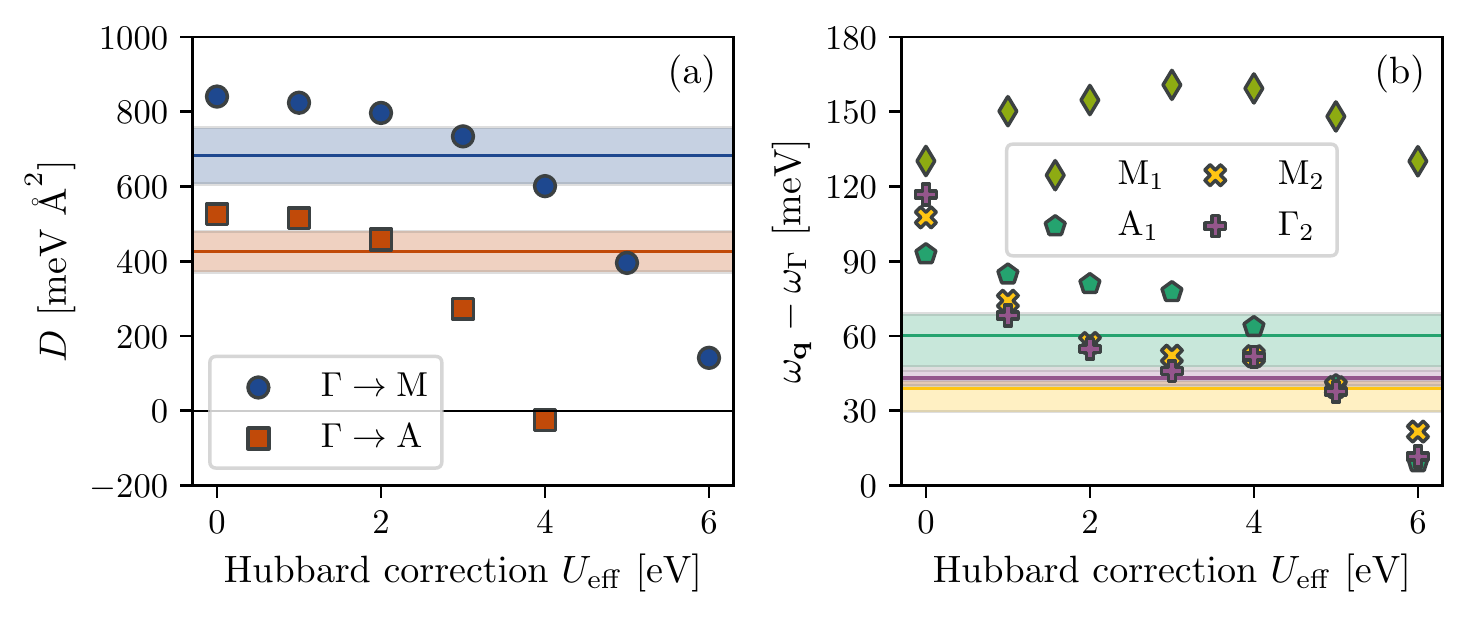}
    \caption{{MnBi magnon dispersion as a function of Hubbard $U_{\mathrm{eff}}$.
    (a) Extracted magnon stiffness along the $\Gamma\rightarrow\mathrm{M}$ and $\Gamma\rightarrow\mathrm{A}$ directions. Horizontal lines and colored areas indicate the experimental best estimates and 1$\sigma$ uncertainty intervals of biquadratic fits to the transverse magnetic structure factor as probed by inelastic neutron scattering \cite{Williams2016}.
    (b) Extracted magnon energies at a selected number of high-symmetry points. The experimentally determined magnon energies are shown by horizontal lines and the colored areas indicate their uncertainties.
    }}
    \label{fig:HubbardU_magnon_dispersion}
\end{figure*}
In Fig. \ref{fig:full_magnon_dispersion_expSpec}, we show the experimental magnon spectrum as measured by inelastic neutron scattering on single-crystaline MnBi \cite{Williams2016} and compare it to the theoretical magnon dispersion relations computed within LR-TDDFT. The magnon dispersion is quite anisotropic, differing substantially between the in-plane ($\Gamma \rightarrow \mathrm{M}$) direction and the out-of-plane ($\Gamma \rightarrow \mathrm{A}$) direction. This difference arises, as the nearest neighbour exchange interaction (which is out-of-plane) is strongly anti-ferromagnetic, while all other couplings are ferromagnetic. In order to fit the magnon spectrum to a Heisenberg model, it is necessary to include interactions up to the 6th nearest neighbours (such that each Mn-site is directly coupled to 40 other Mn sites), and even in this case, the magnon stiffness is underestimated in the model \cite{Williams2016}. The long range of the exchange interactions is attributed the itinerant nature of MnBi and when calculating the magnon dispersion within the ALDA (which is parameter free), we get a better match of the magnon stiffnesses (long range magnon dispersion) to experiment than it is the case with the fitted Heisenberg model. However, the ALDA completely fails to describe the optical magnon mode both quantitatively and qualitatively. In Fig. \ref{fig:full_magnon_dispersion_expSpec}, the optical mode can be observed around the reciprocal lattice vectors $(0, 0, 1)$ and $(1, 0, 1)$, where the ALDA magnon frequencies are more than double compared to experiment. Furthermore, the experimental dispersion attains a minimum at the $\Gamma$-point for the optical mode, most clearly seen at $\mathbf{G}=(1,0,1)$, whereas the ALDA dispersion attains a maximum. This short-coming of the ALDA is to a large extend amended by the inclusion of a Hubbard correction and both the acoustic and optical magnon modes agree reasonably well with the experimental spectrum within the $\lambda$ALDA+U approach with $U_{\mathrm{eff}}=3$ eV. As the optical mode characterizes magnon excitations where the magnetization at neighbouring Mn atoms (out-of-plane) precess out-of-phase, it seems that an appropriate description of static correlation effects is essential for capturing the short-range anti-ferromagnetic exchange interaction. At the same time, it is crucial to incorporate the itinerant nature of MnBi in order to capture the long-range ferromagnetic exchange interactions. This makes the inherent magnetic frustation of MnBi a highly non-trivial problem to treat theoretically. Therefore, it is a noteworthy achievement that we are able to reproduce the experimental magnon dispersion using a simple Hubbard correctional scheme within the framework of LR-TDDFT.

\subsection{Correlation effects and the Hubbard parameter}

So far, we have presented excited state quantities calculated using the effective Hubbard parameter $U_{\mathrm{eff}}=3$ eV. However, the "correct" magnitude of the Hubbard correction cannot be uniquely defined, and in Fig. \ref{fig:HubbardU_magnon_dispersion}(a) we show the magnon stiffness as a function of $U_{\mathrm{eff}}$. We have extracted the experimental magnon stiffness from a biquadratic fit to the inelastic neutron scattering data available \cite{Williams2016}, obtaining values $D_{\mathrm{M}}=683\pm 75\,\mathrm{meV}\,\mathrm{\AA}^2$ and $D_{\mathrm{A}}=426\pm 54\,\mathrm{meV}\,\mathrm{\AA}^2$ for the in-plane and out-of-plane magnon stiffnesses respectively. Using the ALDA kernel, the magnon stiffnesses are slightly overestimated to values of $D_{\mathrm{M}}=841\,\mathrm{meV}\,\mathrm{\AA}^2$ and $D_{\mathrm{A}}=526\,\mathrm{meV}\,\mathrm{\AA}^2$, clearly reproducing the stiffness anisotropy of experiment. Adding an increasing amount of Hubbard correction, as shown in Fig. \ref{fig:HubbardU_magnon_dispersion}(a), the magnon stiffness decreases along both directions, but more so in the out-of-plane direction than it does in-plane. For the $\Gamma \rightarrow \mathrm{A}$ direction, the magnon stiffness reaches negative values for $U_{\mathrm{eff}}\geq 4$ eV, implying that the ferromagnetic ground state becomes dynamically unstable beyond this point, a fact which will be discussed in further detail below. Because the MnBi ground state is indeed ferromagnetic, this sets an upper bound on appropriate values for $U_{\mathrm{eff}}$. Overall, it seems that values of $U_{\mathrm{eff}}\leq 3$ eV provide a reasonable agreement with experiment for the magnon stiffnesses, with the best fit being somewhere in the range of 2-3 eV.

\begin{figure*}[ht]
    \centering
    \includegraphics{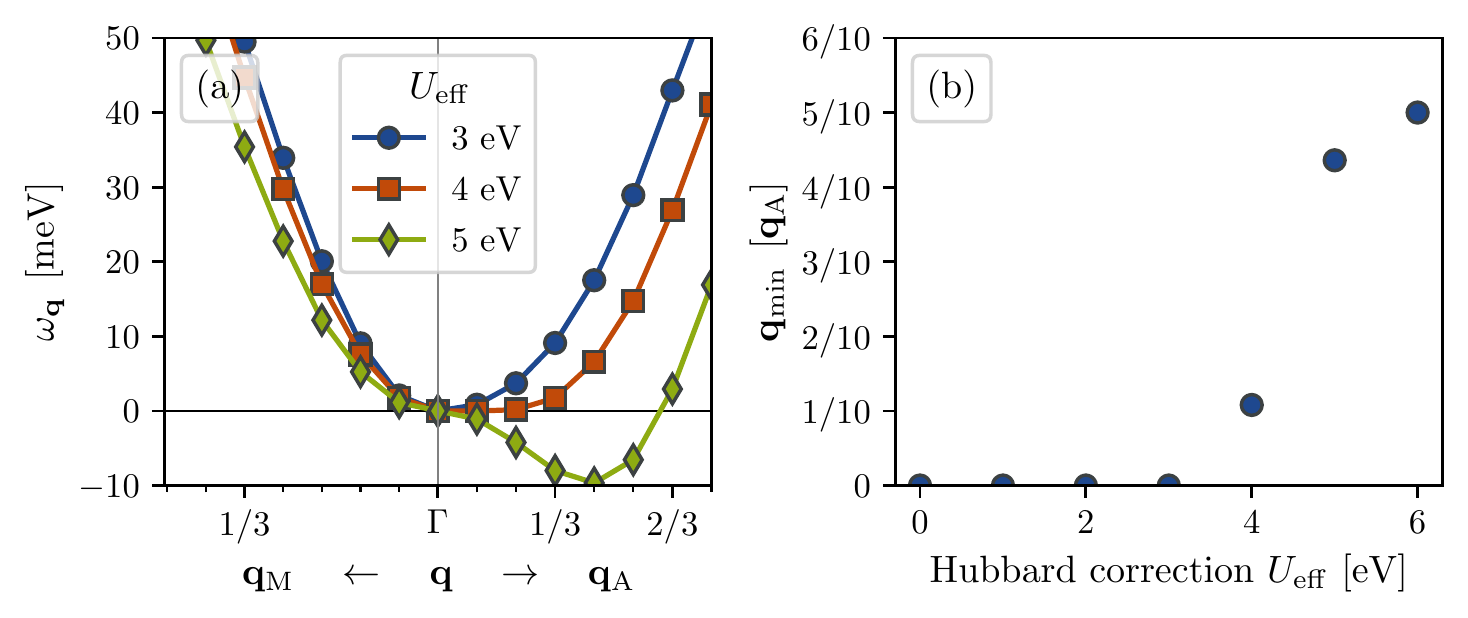}
    \caption{
    (a) Magnon dispersion in the ferromagnetic state at different values for the Hubbard parameter $U_{\mathrm{eff}}$. 
    (b) $\mathbf{q}$-point (in the $\Gamma\rightarrow\mathrm{A}$ direction) where the ferromagnetic magnon dispersion has its minimum as a function of the Hubbard parameter.}
    \label{fig:HubbardU_dynamic_instability}
\end{figure*}
As mentioned previously, it is especially the optical magnon frequencies, for which ALDA falls short. In Fig. \ref{fig:HubbardU_magnon_dispersion}(b) we show the magnon energy at the first BZ M-point and A-point, $(1/2, 0, 0)$ and $(0, 0, 1/2)$ in relative reciprocal coordinates, as well as the second BZ M-point and $\Gamma$-point, $(1/2, 0, 1)$ and $(0, 0, 1)$, as a function of $U_{\mathrm{eff}}$. Using the ALDA kernel, the magnon frequencies at the three high-symmetry points of the second BZ (this includes the A-point) are significantly overestimated. Generally speaking, the three optical magnon frequencies decrease with increasing $U_{\mathrm{eff}}$, coinciding with the experimental reference for separate values in the 3-5 eV range. However, it is not possible to obtain a complete simultaneous match to experiment of both magnon stiffnesses and optical magnon frequencies using the $\lambda$ALDA+U approach. From the results in Fig. \ref{fig:HubbardU_magnon_dispersion}, it seems that in order to reproduce the observed optical magnon frequencies as well as possible, $U_{\mathrm{eff}}$ should be chosen as large as the magnon stiffnesses permit.

Considering all the preceding results, ranging from lattice constants and magnetization to the magnon dispersion relation, it seems that a choice of $U_{\mathrm{eff}}\sim 3$ eV provides the best compromise between various material properties. This finding agrees quite well with the previous efforts of Antropov et al. to determine $U$ and $J$ \cite{Antropov2014}. Using the constrained local spin-density approximation (cLSDA) and constrained random phase approximation (cRPA) methods, they obtain values for $U$ of 4.57 eV and 3.6 eV respectively. As these methods generally tend to overestimate/underestimate the value for $U$, the authors deem a value of $U\sim 4$ eV the most appropriate choice, which along with the cLSDA value for $J=0.97$ eV gives $U_{\mathrm{eff}}=U-J\sim3$ eV.

\begin{figure}[ht]
    \centering
    \includegraphics{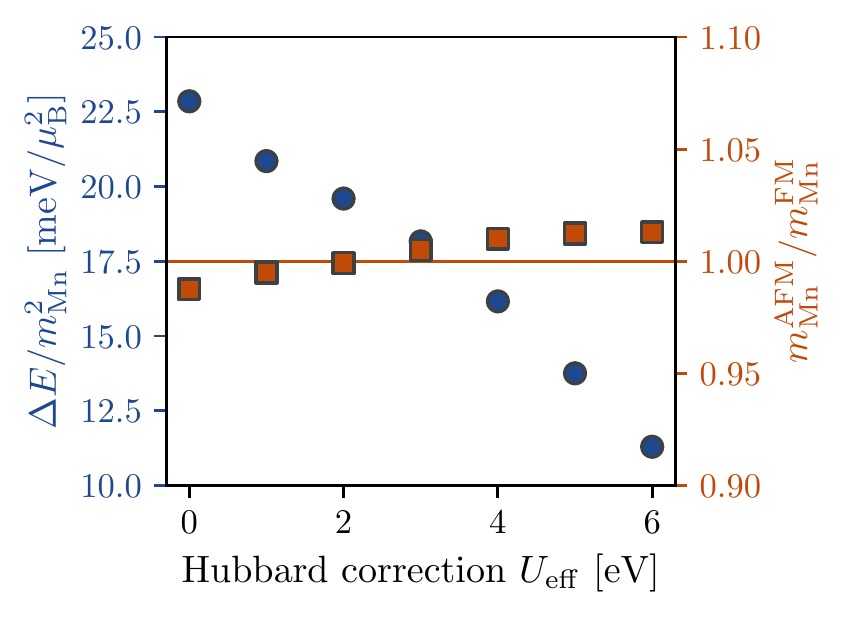}
    \caption{LDA+U effective interlayer exchange coupling (left axis) between the in-plane layers of Mn-atoms, given as a function of $U_{\mathrm{eff}}$. The effective coupling is calculated from the energy difference per magnetic atom $\Delta E = (E_{\mathrm{N\acute{e}el}}-E_{\mathrm{FM}})/2$ and the average local moment on the Mn-atoms $m_{\mathrm{Mn}}=(m_{\mathrm{Mn}}^{\mathrm{FM}}+m_{\mathrm{Mn}}^{\mathrm{N\acute{e}el}})/2$ (within the PAW sphere of radius $r_{\mathrm{c}}=2.1\: a_0$). As a classical Heisenberg model picture of the interlayer coupling relies on the atomic spins being rigid, the relative sizes of the local Mn-moments among the two states is shown on the right axis.}
    \label{fig:HubbardU_interlayer_coupling}
\end{figure}

Finally, we return to the dynamic instability of the ferromagnetic ground state, which occurs for $U_{\mathrm{eff}}\geq 4$ eV. In Fig. \ref{fig:HubbardU_dynamic_instability}(a) we show the magnon dispersion close to the $\Gamma$-point for values of $U_{\mathrm{eff}}$ below, on and above the onset of the instability. At the onset itself ($U_{\mathrm{eff}}\sim 4$ eV), the magnon stiffness vanishes and the dispersion becomes flat along the $\Gamma \rightarrow \mathrm{A}$ direction, meaning that the low-frequency magnons disperse with some power larger than $q^2$ in this direction. Above the onset however, the magnon stiffness is finite and negative, such that the magnon dispersion attains its global minimum away from the $\Gamma$-point and at negative magnon frequencies. This implies that a selection of magnon quasi-particle excitations with finite wave vectors $\mathbf{q}\propto\mathbf{q}_{\mathrm{A}}$ place the system in an energetically more favorable state, than the ferromagnetic starting point, thus rendering the ferromagnetic state dynamically unstable. There is no reason to believe that the correlation effects included in the Hubbard correction quench magnetic order as a whole, rather they seem to enhance the nearest neighbour exchange interaction, which is out-of-plane and antiferromagnetic. 
This is supported by calculations of the energy difference between the ferromagnetic state and the antiferromagnetic state with ferromagnetic alignment in-plane and antiferromagnetic alignment out-of-plane. In Fig. \ref{fig:HubbardU_interlayer_coupling} we illustrate the effective interlayer exchange coupling associated with this energy difference, a coupling which is ferromagnetic, but of decreasing strength with increasing Hubbard correction $U_{\mathrm{eff}}$. It should be stressed that this effective coupling strength is not a valid Heisenberg model parameter in its own right, rather it contains contributions from the exchange couplings at all length scales. 
Beyond the magnetic phase transition $U_{\mathrm{eff}}\geq 4$ eV, the (total) effective interlayer coupling is still ferromagnetic, and it seems likely that the new (hypothetical) ground state would be a helically ordered state. As a first estimate of the helical wave vector, we give in Fig. \ref{fig:HubbardU_dynamic_instability}(b) the wave vector at which the ferromagnetic magnon frequency attains its minimum along the $\Gamma \rightarrow \mathrm{A}$ direction, determined from a fit to the $\lambda$ALDA+U dispersion using gaussian process regression.

\subsection{Hole doping and uniaxial compressive strain}

\begin{figure*}[ht]
    \centering
    \includegraphics{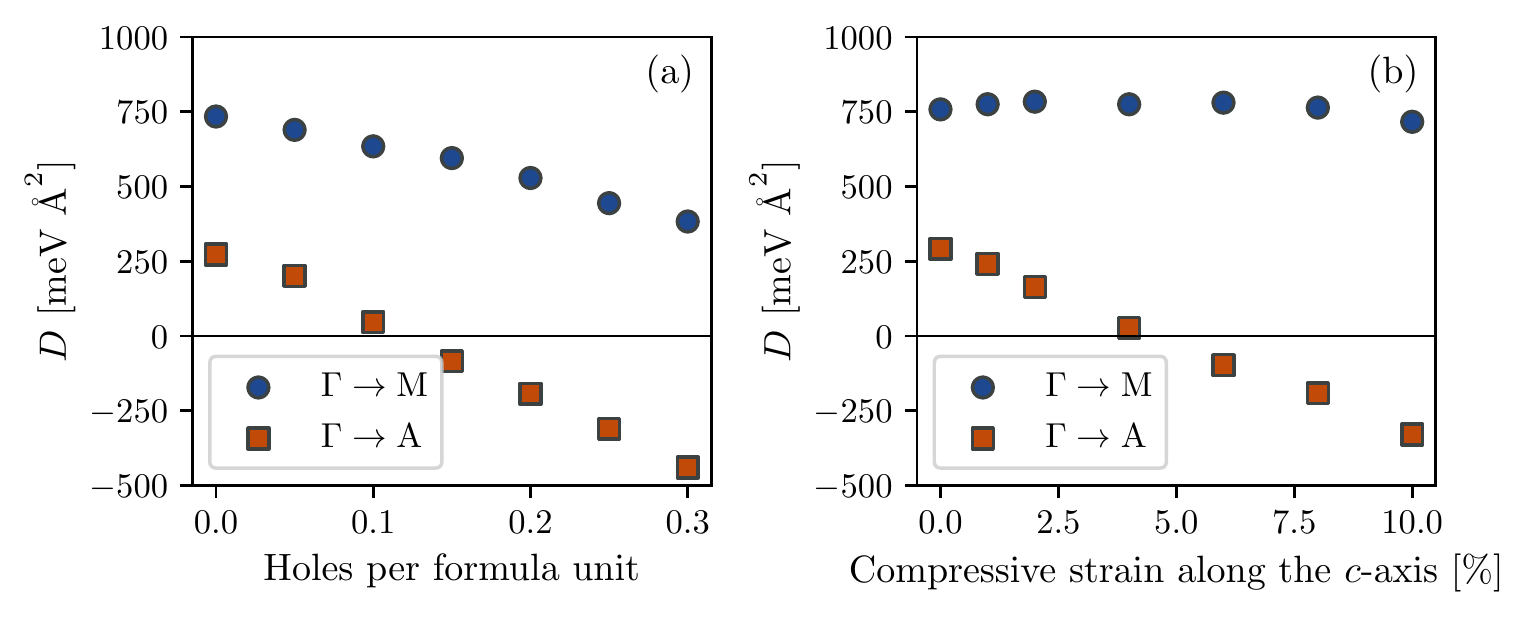}
    \caption{Theoretical magnon stiffness along the $\Gamma\rightarrow\mathrm{M}$ and $\Gamma\rightarrow\mathrm{A}$ directions as 
     (a) a function of removed electrons, 
     (b) a function of uniaxial compressive strain in the $c$-direction. Calculated with the $\lambda$ALDA+U kernel and $U_{\mathrm{eff}}=3$ eV.
    }
    \label{fig:alloying-strain_spin-wave-stiffness}
\end{figure*}
Although the Hubbard parameter $U_{\mathrm{eff}}$ encodes real physical correlations in MnBi, it is not an actual parameter, which can be tuned in experiments. However, as previously hypothesized, the main role of the Hubbard correction regarding a possible phase transition to helical magnetic order seems to be, that it enhances the nearest neighbour anti-ferromagnetic exchange interaction to such an extent, that the ferromagnetic state becomes dynamically unstable. Whereas we cannot tune $U_{\mathrm{eff}}$ in real life, we can try to enhance the anti-ferromagnetic exchange and possibly realize a magnetic phase transition to helical order in this way. To this end, we investigate two different approaches: Hole doping and uniaxial compressive strain. Investigating the effect of hole doping is motivated by the fact that Mn sits to the right of Cr in the periodic table. Cr is well-known for exhibiting strong anti-ferromagnetic exchange interactions and by substituting a small amount of Mn with Cr it may be possible to induce a phase transition into a helically ordered ground state. Because large supercell LR-TDDFT computations currently are out-of-scope, we simulate this scenario by introducing holes into the electronic structure of MnBi (moving the Fermi level down). We do this with $U_{\mathrm{eff}}$ fixed to a value of 3 eV. The motivation for investigating strain effects on the magnon dynamics is more straightforward. If we can compress the nearest neighbour bond length, the anti-ferromagnetic exchange interaction should be enhanced. However, if we apply a hydrostatic pressure and reduce the crystal volume from all directions, also the ferromagnetic exchange interactions are expected to increase in strength. Thus, to simplify the picture, we compress the crystal along the $c$-axis, while keeping the volume constant through an expansion of the in-plane lattice constant. In experiment, this would to some extent correspond to applying a uniaxial pressure.

\begin{figure*}[ht]
    \centering
    \includegraphics{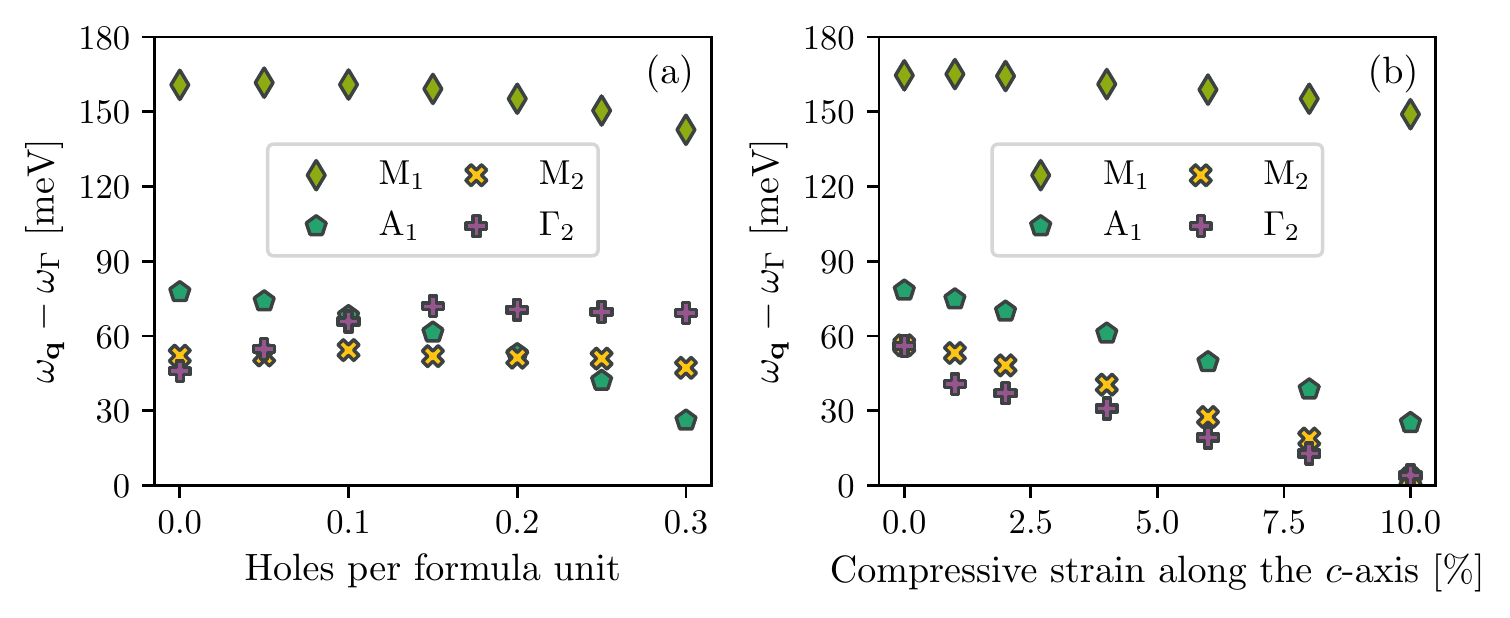}
    \caption{Theoretical magnon energies at a selected number of high-symmetry points as 
    (a) a function of removed electrons, 
    (b) a function of uniaxial compressive strain in the $c$-direction. Calculated with the $\lambda$ALDA+U kernel and $U_{\mathrm{eff}}=3$ eV.
    }
    \label{fig:alloying-strain_bandwidth}
\end{figure*}
In Fig. \ref{fig:alloying-strain_spin-wave-stiffness}, we show the magnon stiffness as a function of hole doping and uniaxial compressive strain. In both cases, the magnon stiffness decreases more or less linearly along the $\Gamma \rightarrow \mathrm{A}$ direction and the ferromagnetic state becomes unstable at 0.12 holes per formula unit and 4.4\% compressive strain along the $c$-axis (determined from linear interpolation). When hole-doping the system, also the in-plane magnon stiffness along the $\Gamma \rightarrow \mathrm{M}$ direction decreases. This fits well with the intention of simulating a substitution of Mn with Cr, as one would also expect a weakening of the long-range ferromagnetic exchange interactions in this case. For the uniaxial compressive strain, it is less clear what to expect for the in-plane magnon stiffness. With the expansion of the in-plane lattice constant, the in-plane ferromagnetic exchange interactions are expected to become weaker, leading to a reduction in the frequency scale for $D_{\mathrm{M}}$. At the same time however, the length scale for $D_{\mathrm{M}}$ increases with an expanding $a$ lattice constant.
Seemingly, these opposite sided effects cancel out, as the in-plane magnon stiffness in Fig. \ref{fig:alloying-strain_spin-wave-stiffness}(b) is seen to be largely unaffected by the uniaxial compressive strain, at least for strains below 10\% (which are really 5\% strains in-plane). 
The weakening of in-plane ferromagnetic exchange interactions is more clear from the in-plane magnon band width (i.e. the magnon frequency at the first BZ M-point), for which there is a decrease with \textit{both} hole doping and uniaxial compressive strain, as seen in Fig. \ref{fig:alloying-strain_bandwidth}. 
In addition, also the magnon frequency in the A-point decreases with both hole-doping and uniaxial compressive strain. However, this is not the case for the second BZ M-point and $\Gamma$-point for which the magnon frequencies only decrease with strain. On the contrary, the magnon frequency in the second BZ center actually increases with hole doping. This is somewhat of a surprise, since the second BZ $\Gamma$-point magnons correspond to spin wave excitations where the nearest neighbour Mn atoms acquire opposite phases. Once again, this emphasizes that the magnetic frustration in MnBi is a highly non-trivial problem, due to the importance of static correlation effects and the fact that the long-range ferromagnetic exchange interactions cannot be boiled down to a simple set of short-range couplings.

Experimental studies have shown that it is indeed possible to dope MnBi with Cr, at least for thin films \cite{Bandaru1998,Bandaru1999} and melt-spun ribbons \cite{Anand2019}. It may even be an advantage for the synthesis to add Cr, as it helps to stabilize the formation of ferromagnetic MnBi as opposed to a decomposition into paramagnetic Mn$_{1.08}$Bi and Bi \cite{Anand2019}. For both thin films and ribbons, the Curie temperatures of the investigated Mn$_{1-x}$Cr$_{x}$Bi alloys lie below the segregation temperature of 630 K. Based on these studies, it is not completely clear how the Curie temperature depends on Cr doping levels, since the samples are rather inhomogeneous. As an example, the Cr content of the thin films was shown to depend on depth, with Cr concentrated at the surface \cite{Bandaru1999}. With 10\% Cr in the overall composition (comparable to the critical doping level for the helical phase transition according to the results of Fig. \ref{fig:alloying-strain_spin-wave-stiffness}(a)), the Curie temperature was decreased to 523 K and 546 K for the thin films and ribbons respectively. The fact that Cr doping lowers the Curie temperature is in good agreement with our results, but a confirmation of the calculated trends and existence of a magnetic phase transition (possibly at low temperatures) requires further experimental studies.

\begin{figure*}[ht]
    \centering
    \includegraphics{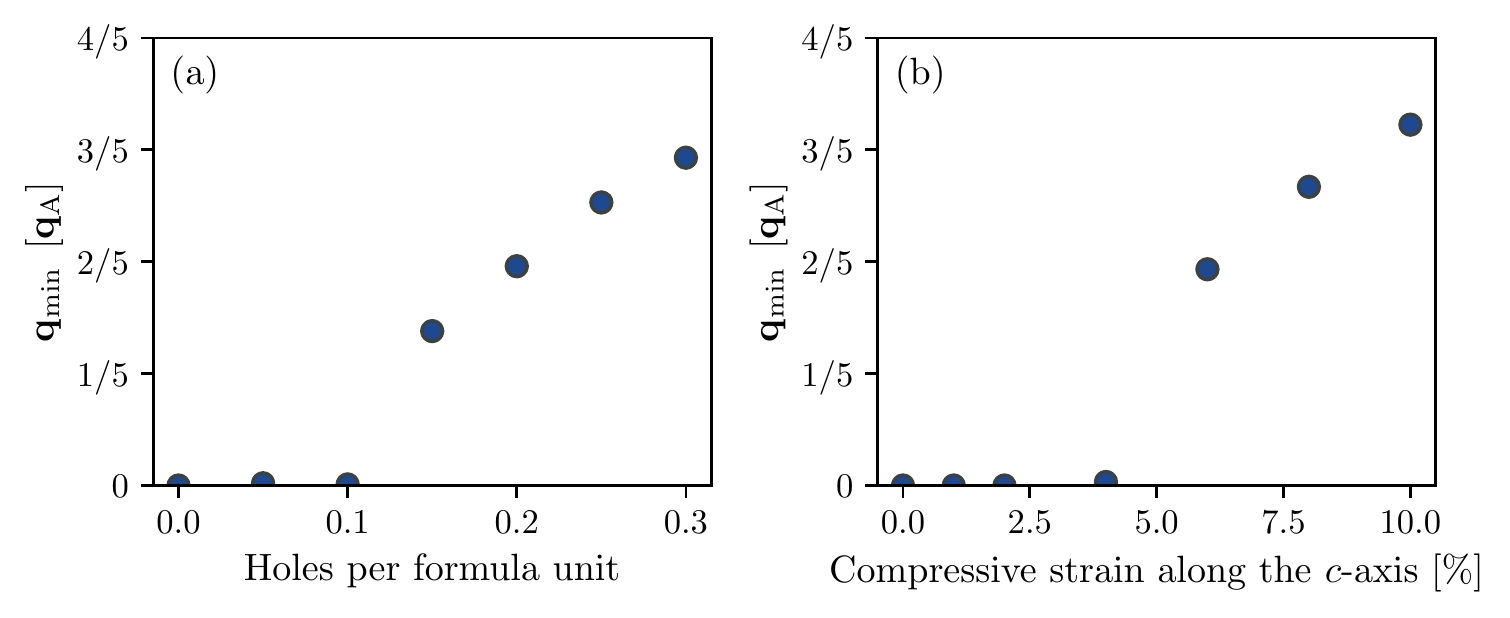}
    \caption{$\mathbf{q}$-point where the ferromagnetic magnon dispersion has its minimum along the $\Gamma\rightarrow\mathrm{A}$ path as
    (a) a function of removed electrons, 
    (b) a function of uniaxial compressive strain in the $c$-direction. Calculated with the $\lambda$ALDA+U kernel and $U_{\mathrm{eff}}=3$ eV.
    }
    \label{fig:alloying-strain_qmin}
\end{figure*}
To our knowledge, it is an open question, how MnBi behaves under uniaxial pressure. From the present theoretical investigations, it does not seem realistic to induce a phase transition to helical magnetic order in a uniaxial pressure cell, as one would expect the crystal to break long before obtaining a 4.4\% uniaxial compressive strain. However, based on the fact that $D_{\mathrm{A}}$ decreases linearly with both hole-doping and strain, as seen in Fig. \ref{fig:alloying-strain_spin-wave-stiffness}, one could hope to combine the two effects, such that the hypothetical helical phase transition of the Mn$_{1-x}$Cr$_{x}$Bi alloys could be induced by pressure for Cr-doping levels close to the phase transition. Indeed, it seems that both effects produce similar trends for the minimal frequency magnon wave vectors beyond the magnetic phase transition, as seen in Fig. \ref{fig:alloying-strain_qmin}.

\section{Discussion}\label{sec:discussion}

In this paper, we have mainly discussed the magnon dynamics of pristine MnBi and Cr doped alloys. However, the results can be regarded as an initial step in a broader context, since many of the quantum magnetic phenomena studied here are shared by the family of transition metal pnictides. Among the Mn and Cr based compounds, helical magnetic order is quite common and the helical wave vectors have also previously been shown to depend on the relative concentrations of Mn and Cr \cite{Wang2016}. As an example, the Mn$_{1-x}$Cr$_{x}$As phase diagram includes a phase transition from ferromagnetic MnAs to helically ordered CrAs \cite{Luo2017}. In addition, MnP has been shown to exhibit pressure induced magnetic phase transitions between ferromagnetic and helical order as well as pressure induced super-conductivity \cite{Cheng2015,Wang2016}, which is also found in CrAs \cite{Wu2014}.

Also the inherent magnetic frustration in the transition metal pnictide family has been the subject of several previous studies. Based on analysis of the band structure and partial occupation of orbitals in MnP, Goodeneough provided a qualitative explanation for the existence of both anti-ferromagnetic and ferromagnetic order at different temperatures \cite{Goodenough1964}. He argued that the half-filling of the localized Mn $t_{2g}$ orbital directed towards the nearest Mn neighbours calls for an anti-ferromagnetic ordering while the three-fourths filling of the collective, yet narrow, 3$d$ bands representing the remaining $t_{2g}$ orbitals calls for ferromagnetic ordering. Similar considerations have also been explored for MnAs, where exchange striction also plays a role in the competition between the NiAs and MnP structural phases \cite{Goodenough1967}.

In this context, the theoretical prediction that magnetic order in MnBi is characterized by a competition between nearest neighbour anti-ferromagnetic exchange and long-range ferromagnetic exchange has some precedent, as similar effects have been observed in closely related members of the transition metal pnictide family. Similarly, it seems likely that the prediction of a phase transition to helical magnetic order in Cr doped MnBi could be correct and that the phase transition would be strain sensitive.

The coupling between structural and magnetic degrees of freedom may also help to explain, why the Hubbard correction has such a strong influence on the out-of-plane lattice constant $c$, as show in Fig. \ref{fig:ground_state_properties}(a). As the appropriate correlation effects are included with increasing $U_{\mathrm{eff}}$ and the anti-ferromagnetic exchange interaction between nearest neighbours is increased, it becomes favorable to increase the nearest neighbour distance as a compensation for the ground state remaining ferromagnetic. Similarly, a magnetic phase transition to helical order could be accompanied by a structural compression in the $c$-direction, as the anti-ferromagnetic nearest neighbour exchange interaction does no longer need to be compensated to the same extent. Another possibility is that MnBi undergoes a structural phase transition from the hexagonal NiAs structure into the orthorhombic MnP structure with Cr doping, as it is the case in the Mn$_{1-x}$Cr$_{x}$As phase diagram.

Although the occurrence of similar phenomena in other transition metal pnictides inspire confidence in the qualitative predictions made on the basis of this study, the quantitative predictions may depend somewhat on the details in the theoretical representation of doping and strain. The by-hand introduction of hole-doping and uniaxial compressive strain should be viewed as efforts to study the underlying physical mechanisms and not to provide accurate estimates for e.g. the critical Cr doping level for the magnetic phase transition. Actually, the critical doping levels and strains for the phase transition are likely to be underestimated, as the Hubbard parameter $U_{\mathrm{eff}}=3$ eV used here leads to an underestimate of the out-of-plane magnon stiffness $D_{\mathrm{A}}$, as seen in Fig. \ref{fig:HubbardU_magnon_dispersion}(a). In this sense, the undoped an unstrained MnBi is closer to a helical phase transition in our simulations, than the experiments predict (based on the magnon stiffness).

\section{Conclusion}\label{sec:conclusion}

MnBi has a long history of experimental as well as theoretical investigations based on its attractive properties for technological applications. As we have shown in this study, MnBi exhibits a non-trivial inherent magnetic frustration, which makes it an intriguing subject for theoretical studies, but also implies a potential for future discovery of new magnetic phases. The nearest neighbour exchange interactions between Mn 3$d$ electrons are strongly anti-ferromagnetic and highly susceptible to correlation effects, but despite the strength of these interactions, the ground state magnetic order is determined by the long-range ferromagnetic exchange. Because the competing magnetic interactions arise from electrons of localized and itinerant character respectively, it is a substantial theoretical challenge to provide an appropriate description of the magnetic frustration in MnBi.

In this study, we have shown that it is in fact possible to capture the magnetic frustration from the perspective of (LR-TD)DFT calculations, both for the ground state properties of MnBi, but also for the magnon dynamics. The itinerant and localized correlations in MnBi have been described at the LDA+U level and for the LR-TDDFT calculations, we utilize a scalar rescaling of the ALDA kernel based on considerations of the Goldstone criterion for the homogeneous electron gas. With the rescaled $\lambda$ALDA+U kernel, we are able to reproduce the experimental magnon dispersion using a Hubbard correction of $U_{\mathrm{eff}}=3$ eV, which in turn also provides ground state properties that are in accordance with experiment. In this way, the present study may pave the way for future theoretical studies of magnon dynamics in the transition metal pnictide family. To this end, we hope that first principles calculations can provide a new angle of insight into the wide range of phenomena driven by magnetic frustration and magnetic fluctuations.

With an \textit{ab initio} description of the magnon dynamics of MnBi in place, we have explored some of the phenomena that arise from the magnetic frustration. We have shown that an increase of the local electronic correlations gives rise to a decrease in the magnon stiffness out-of-plane ($D_{\mathrm{A}}$) due to an increased strength of the anti-ferromagnetic exchange interactions between the nearest neighbours. At $U_{\mathrm{eff}}\sim 4$ eV, the magnon stiffness changes sign, meaning that a phase transition takes place, most likely in the favor of a phase of helical order. Whereas the increase in electronic correlations is an artificial one, we have shown that a similar increase in anti-ferromagnetic interaction strength may be imposed using hole-doping or uniaxial compressive strain, in both cases leading to a similar phase transition. In particular, it seems realistic to realize this phase transition by substituting 10-20\% of the Mn content with Cr. Similarly, it may be possible to induce the transition by applying uniaxial pressure, but only for compositions of Mn$_{1-x}$Cr$_{x}$Bi close to the critical Cr doping level. Furthermore, we predict the helical wave vector $\mathbf{q}$ to be sensitive to the Cr doping level and strain, especially for doping levels and strains close to the phase transition.

To further unravel the inherent magnetic frustration in MnBi, additional experiments on single-crystaline MnBi are highly desirable. Especially investigations of the effect of Cr doping and a possible phase transition to helical magnetic order are of great interest. Such investigations could potentially provide an improved understanding of the physical mechanisms underlying a wide range of members in the family of transition metal pnictides.

\end{document}